\documentclass[fleqn,usenatbib]{mnras}

\usepackage{newtxtext,newtxmath}
\usepackage[T1]{fontenc}

\DeclareRobustCommand{\VAN}[3]{#2}
\let\VANthebibliography\thebibliography
\def\thebibliography{\DeclareRobustCommand{\VAN}[3]{##3}\VANthebibliography}

\usepackage{graphicx}	% Including figure files
\usepackage{amsmath}	% Advanced maths commands
\usepackage{multirow}
\usepackage{subcaption}
\usepackage[normalem]{ulem}
\usepackage{xcolor}

\title[Phase-space simulations of prompt cusps]{Phase-space simulations of prompt cusps: simulating the formation of the first haloes without artificial fragmentation}

\author[Ondaro-Mallea et al.]{
Lurdes Ondaro-Mallea$^{1}$\thanks{E-mail: lurdes.ondaro@dipc.org}, Raul E. Angulo$^{1,2}$, Jens Stücker$^{1}$, Oliver Hahn$^{3,4}$, Simon D.M. White$^{5}$.
\\
$^{1}$Donostia International Physics Center, Manuel Lardizabal Ibilbidea, 4, 20018 Donostia, Gipuzkoa\\
$^{2}$IKERBASQUE, Basque Foundation for Science, 48013, Bilbao, Spain\\
$^{3}$Department of Astrophysics, University of Vienna, T\"urkenschanzstra\ss e 17, 1180 Vienna, Austria.\\
$^{4}$Department of Mathematics, University of Vienna, Oskar-Morgenstern-Platz 1, 1090 Vienna, Austria.\\
$^{5}$Max-Planck-Institut f\"{u}r Astrophysik, Karl-Schwarzschild-Str. 1, 85748, Garching, Germany
}
\date{Accepted XXX. Received YYY; in original form ZZZ}
\pubyear{2023}
\begin{document}
\label{firstpage}
\pagerange{\pageref{firstpage}--\pageref{lastpage}}
\maketitle
\begin{abstract}
 The first generation of haloes forms from the collapse of the smallest peaks in the initial density field. $N$-body simulations of this process suggest a prompt formation of a steep power-law cusp, but these calculations are plagued by numerical artifacts which casts some doubt on this result. Here, we develop new simulation methods based on the dark matter phase-space sheet approach and present results which are entirely free of artificial clumps. We find that a cusp with density $\rho\propto r^{-1.5}$ is indeed formed promptly, subsequently accreting a more extended halo and participating in the hierarchical growth of later halo generations. However, our simulations also suggest that the presence of artificial clumps just before peak collapse can significantly shallow the inner profiles of the cusps. We use $N$-body simulations with controlled amounts of small-scale power to place a conservative upper limit on the scales affected by artificial clumps. Finally, we used these results to simulate the collapse of the first generation of peaks of various types and in different cosmologies, finding prompt cusps to form in all cases. We conclude that prompt cusps are a generic feature of the collapse of peaks on the free-streaming scale of the initial density field, and their structure can safely be studied using $N$-body simulations provided care is taken to excise the region potentially affected by artificial clumps.
\end{abstract}
%%%%%%%%%%%%%%%%%%%%%%%%%%%%%%%%%%%%%%%%%%%%%%%%%%

%%%%%%%%%%%%%%%%% BODY OF PAPER %%%%%%%%%%%%%%%%%%

\section{Introduction}

$N$-body simulations have shown that the spherically averaged density profile of dark matter haloes within their virial radii is well described by the remarkably simple functional form  
\begin{equation}
 \rho(r) = \frac{\rho_sr_s^3}{r(r+r_s)^2},
% \rho(r) = \frac{\rho_s}{r/r_s(r/r_s+1)^2},
\end{equation}

\noindent with just two parameters ($\rho_s$ and $r_s$), the so-called NFW profile \citep{Navarro:1996,Navarro:1997}.

In the inner parts of the halo, this reduces to a density profile proportional to $r^{-1}$, while in the outer parts it drops like $r^{-3}$. Later studies have found that a weak variation in shape depending on halo mass and formation history can further improve the fit to simulated profiles \citep{Navarro:2004,Ludlow:2013,Lopez:2022}. Furthermore,  at radii close to and beyond the virial radius, density profiles are sensitive to the recent mass accretion rate \citep{Diemer:2014,Lucie-Smith:2022}, and the modelling needs to be extended, adding new parameters \citep{Diemer:2022a,Diemer:2022b}. Nonetheless, the overall trends described by NFW remain valid. In particular, their profile provides a reasonable fit over more than three orders of magnitude in radius \citep{Springel:2008} and 20 orders of magnitude in mass \citep{Wang:2020}. 

This picture changes when simulations resolve the free-streaming scale of dark matter. Small-scale perturbations in the dark matter field are washed out by the velocity dispersion of dark matter, imprinting a cut-off in the linear power spectrum. This sets a minimum scale for the perturbations that grow under the action of gravity and eventually collapse to form the first haloes. $N$-body simulations of this initial phase predict prompt formation of a structure that differs from standard CDM haloes, a power-law cusp, $\rho \propto r^{-1.5}$ \citep{Diemand:2005,Ishiyama:2010,Anderhalden:2013,Ishiyama:2014,Polisensky:2015,Angulo:2017,Ogiya:2018,Colombi:2021,Delos:2022}. 

In contrast, studies of present-day haloes of mass close to the free-streaming mass have found profiles in overall agreement with NFW \citep{Lovell:2014,Wang:2020}. There is consensus that after growing several orders of magnitude in mass, both via smooth accretion and major mergers, first haloes converge towards NFW form over the bulk of their mass \citep{Lovell:2014,Angulo:2017,Wang:2020,Delos:2022}. The survival of prompt cusps in the innermost region of the halo at the end of this process is still debated: some papers claim that they are destroyed by major mergers \citep{Angulo:2017,Ogiya:2016}, while others find that they are largely unaffected \cite{Delos:2022}. 

Analytic models of halo formation have so far not succeeded in predicting the universal NFW form. Early spherical models for self-similar infall  predicted steep cusps $\rho \propto r^{-\gamma}$ with $\gamma \geq 2.0$ \citep{Gunn:1972,Fillmore:1984,Bertschinger:1985}. This lower limit is a consequence of the assumption of purely radial orbits, and smaller values of $\gamma$ are possible if the assumption of strict spherical symmetry is relaxed \citep{White:1992, Ryden:1993,Vogelsberger:2011,Lithwick:2011}. Recently, \cite{White:2022} presented a self-similar model  applicable to the moments immediately after the initial collapse of a smooth density peak, predicting $\rho \propto r^{-1.7}$, which is close to but still in tension with $N$-body predictions for the structure of first haloes. A short-coming of many of these models is the lack of a consistent treatment of anisotropic collapse, tidal fields and the build-up of angular momentum.

Thus, the problem of structure formation is not yet completely solved, and several open questions remain. On the one hand, a first principle explanation of the emergence of NFW profiles and their link with the prompt cusps of the very first haloes is still a matter of debate. Even if there are strong arguments that point to the role of mass accretion histories \citep{Ludlow:2013}, smooth accretion \citep{Delos:2022} or even mergers \citep{Ogiya:2016,Angulo:2017} as important ingredients determining evolution towards the universal NFW profile, a complete physical picture is still missing. On the other hand, the validity of the $N$-body predictions should not be taken for granted. In the context of CDM haloes, extensive convergence studies have been made in order to constrain the effects of the collisionality introduced by the $N$-body technique on the profiles \citep{Power:2003, Ludlow:2019}. Nevertheless, there remains, perhaps, a slight possibility that the profiles have, by a conspiracy of effects, converged to the wrong result. In particular, it has been suggested that the NFW profile could emerge as the result of spurious relaxation due to spurious particle scattering effects in $N$-body simulations \citep{Baushev:2015}.

When dealing with the formation of the first haloes, a different discreteness effect comes into play. $N$-body simulations of warm dark matter (WDM) -- i.e. cosmologies with a relatively large small-scale cut-off in the power spectrum -- are known to create artificial clumps due to discreteness noise \citep{Wang:2007,Power:2016}. The anisotropic collapse of dark matter forms filaments that are collapsed in two directions, and almost completely cold in the third one. The absence of velocity dispersion in the third direction (i.e. along the filament direction) makes filaments sensitive to discreteness fluctuations in the density field. These can seed an anisotropic Jeans-like instability that leads to the formation of artificial clumps along the filament. % inaccuracies in the gravity solver and facilitates artificial fragmentation when numerical errors add up coherently between subsequent time steps.
Furthermore, the density estimate provided by $N$-body simulations is a noisy approximation of the underlying continuous density field, especially in low density regions \citep{Abel:2012, Shandarin:2012} such as filaments, making them overly likely to fragment.

The formation of the first haloes in $N$-body simulations may thus be affected by the presence of artificial fragments in pre-collapse structures. This might alter the resulting profile in some way, thus rendering the predictions of $N$-body simulations unreliable. In previous work, this problem has been minimized by increasing particle numbers in the simulations, so that the fragments are pushed to smaller scales, and hopefully, their effects too \citep{Ishiyama:2014, Delos:2022}. 

The only convincing way to validate $N$-body predictions in the regime of first halo formation is to use an independent numerical scheme that does not suffer from discreteness and artificial fragmentation. In this paper, we use sheet-based numerical techniques to do just this \citep{Hahn:2013,Hahn:2016,Sousbie:2016,Stucker:2020}. By comparing our results with standard $N$-body simulations, we can test whether discreteness artifacts leave an imprint on the structure of the first haloes. If they do not, $N$-body simulations can safely be used to study these objects. Otherwise, it is important to be aware of the limitations of $N$-body simulations in this context. 

This paper is structured as follows. In Section \ref{sec:numerics}, we introduce the numerical techniques employed in our simulations, and in Section \ref{sec:simulations} we describe the specific set-up needed here. In Section \ref{sec:halos}, we present simulations of the formation of three haloes that are free of any artificial fragmentation, comparing them to $N$-body simulations from identical initial conditions. In Section \ref{sec:artfrags}, we make use of a controlled set of $N$-body simulations to mimic the effect of artificial fragmentation. In the following Section, we use insights from this exercise to set limits on the range of validity of $N$-body simulations. Finally, in Section \ref{sec:universality} we use a large set of $N$-body simulations to study the universality of the inner structure of the first haloes as a function of the warmth of dark matter, as well as the effect of varying the shape of the cut-off of the power spectrum and formation redshift. We conclude in Section \ref{sec:concl}.

\section{Numerical techniques}\label{sec:numerics}

\subsection{Dark matter sheet}\label{sec:dmsheet}

Most particle physics models for dark matter predict that it has a non-zero velocity dispersion at the time it decouples from the other components of the Universe (e.g. \cite{Arcadi:2018,Adhikari:2017}). Its weak subsequent interactions with itself or with anything else allow dark matter particles to free-stream, washing out perturbations smaller than an effective free-streaming distance \citep{Cyr-Racine:2016,Murgia:2017,Niemeyer:2020}. This sets a minimum size or mass for structures in the late Universe. In the absence of such initial velocity dispersion, perturbations could collapse and form structures down to arbitrarily small scales.

After decoupling, the distribution of dark particles must thus be specified in 6 dimensions -- three for position and three for velocity. During the matter dominated era before the start of nonlinear structure formation, the dark matter velocity dispersion (which decays as $1/a$ once nonrelativistic) drops to values much smaller than needed to traverse the free-streaming length in a Hubble time. 

 As a result, the residual dispersion has no significant effect on the collapse of the first objects and can safely be neglected, although it does determine the maximum possible central dark matter density in all nonlinear objects \citep{Tremaine:1979}. Thus, during nonlinear structure formation, dark matter can be assumed to occupy a 3-dimensional submanifold of the full 6-dimensional phase-space, which is often referred to as the dark matter sheet \citep{Abel:2012,Shandarin:2012}. Gravitational evolution stretches, bends and folds this sheet, but it is never torn and it conserves mass, properties that are exploited by sheet-based cosmological simulations.

\subsection{Sheet-based simulations}\label{sec:sheet_based_sims}

Sheet-based simulation schemes \citep{Hahn:2013,Hahn:2016,Sousbie:2016} provide an alternative to $N$-body methods when solving the equations for the gravitational evolution of dark matter. They tessellate the dark matter sheet in the initial conditions and follow the evolution of each element in time. This is done by evolving the vertices of the elements (flow tracers) according to the gravitational acceleration they experience. As the connectivity of the dark matter sheet is conserved and known, one can interpolate between the vertices of the phase-space elements and recover a continuous description of the dark matter sheet at any moment.

Notice that despite using a different discretization technique, $N$-body codes also follow the evolution of dark matter in phase-space. A key distinction, however, is that the mass is deposited throughout each phase-space element in sheet-based schemes, rather than being concentrated in point particles. As a result, sheet-based density estimates are much smoother than the estimates obtained from $N$-body methods. This greatly reduces discreteness effects and almost eliminates the problem of artificial fragmentation.

These advantages come at the cost of computationally expensive calculations, in particular, inside haloes, where rapid mixing takes place and the phase-space sheet becomes extremely convoluted. In order to retain a good, unbiased description, it would be necessary to refine the phase-space elements recursively to a level which is, in practice, prohibitive  \citep{Colombi:2021}. For these reasons, we adopt the 'sheet and release' (S+R) hybrid scheme presented in \cite{Stucker:2020}, where the interiors of haloes are simulated with $N$-body techniques, while a sheet-based method is used everywhere else. Throughout this paper, we will refer to sheet-based and S+R techniques interchangeably.

\subsection{Refinement before the collapse}\label{sec:refinement}

\begin{figure}
    \centering
    \includegraphics[width=0.7\columnwidth]{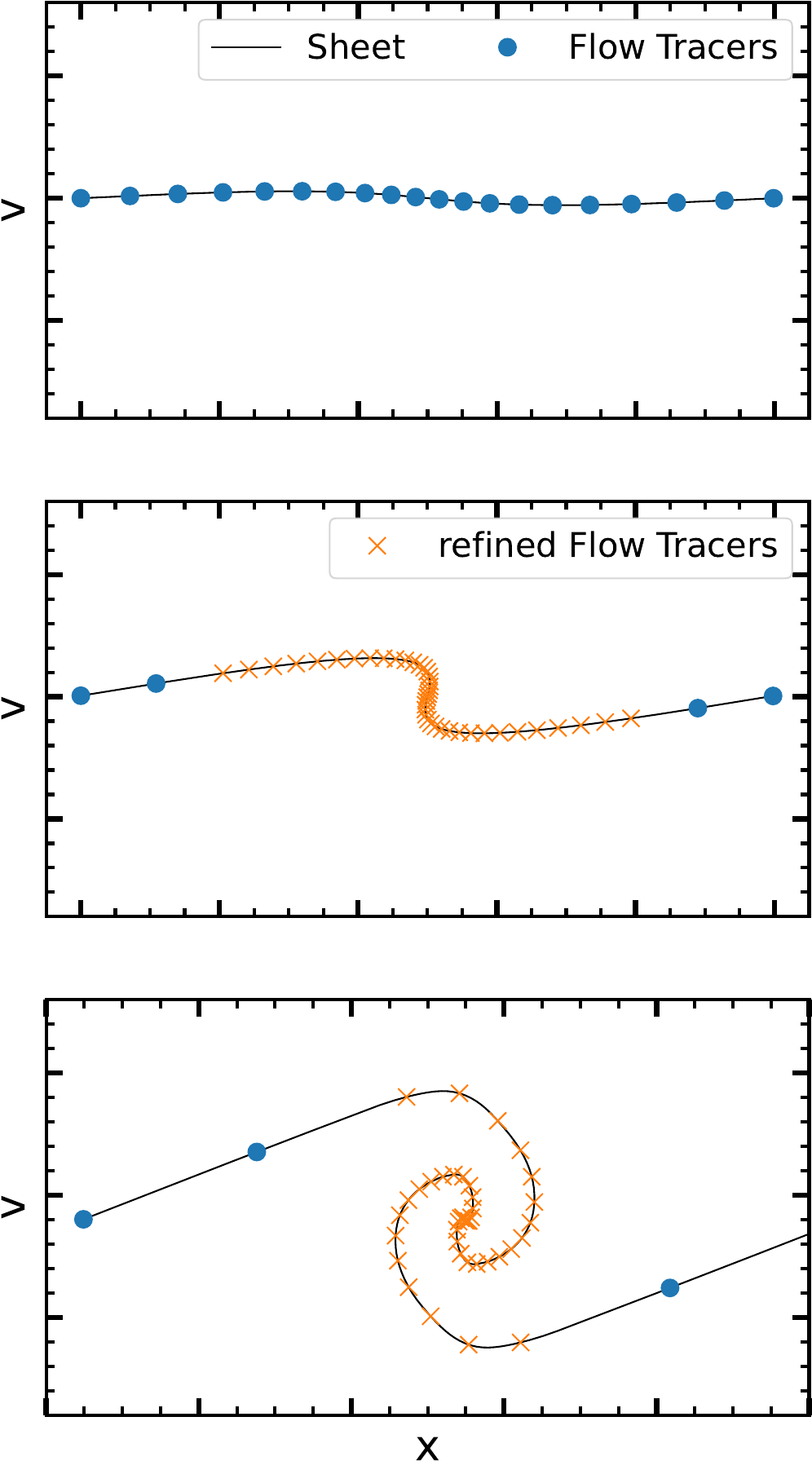}
    \caption{Phase-space evolution of a 1D dark matter sheet, explaining our refinement technique. In the initial conditions (upper panel), the dark matter sheet is discretized with relatively few flow tracers. Soon before the collapse (middle panel), we refine the region of interest, increasing the number of flow tracers. Once the halo has formed (lower panel), it is described at high-resolution, while the rest of the sheet keeps its original resolution. In 3D the sheet rapidly becomes highly convoluted in the halo interior so we switch to an $N$-body description in which the flow tracers are converted into mass-bearing $N$-body particles.}
    \label{fig:diagram_sheet}
\end{figure}

Simulating the dark matter sheet to the point of halo formation would be very expensive if carried out entirely at high resolution, so we introduce a multi-scale refinement technique. We refine regions of interest -- for instance, where haloes will form -- just before halo formation, rather than in the initial conditions. Thus, this scheme can be thought as a zoom-in in the quasi-linear regime, instead of a traditional zoom-in technique applied in the linear initial conditions.  We study its convergence in Appendix \ref{sec:conv_sheet}.

In Figure \ref{fig:diagram_sheet} we illustrate the main features of our method using the phase-space evolution of a 1D dark matter sheet. In the initial conditions (upper panel), the dark matter sheet has a simple shape in phase space, and a small number of phase-space elements are sufficient to describe it adequately -- i.e. interpolation between  flow tracers provides an accurate description of the underlying continuous sheet. As it undergoes gravitational evolution, the sheet becomes more complex, and the phase-space elements get distorted (middle panel). By the equivalent time in 3D, filaments have formed, and the halo is about to collapse, so we are already in the quasi-linear regime. The low-resolution description of the sheet is nevertheless still accurate. Taking advantage of this, we refine the region that will collapse further by inserting  more flow-tracers, keeping other regions at the original resolution. Notice that this possibility is unique to sheet-based methods and could not be implemented in $N$-body schemes because it requires an accurate continuous description of the underlying phase-space distribution. Thus halo collapse is resolved at high resolution (lower panel). In 3D the sheet soon becomes too convoluted to follow any further, and we then convert the flow tracers to mass-bearing $N$-body particles (S+R).

This approach has several advantages. Firstly, it makes the algorithm more robust to numerical instabilities so filaments are less likely to fragment. Secondly, it greatly lowers the computational cost of running the simulation. In addition, one can refine regions of interest as many times as possible given computing time constraints, thereby increasing the time over which first halo collapse can be followed before conversion from sheet-based to $N$-body methods is required. \footnote{Notice that this refinement technique is only applicable when no real physical structure is expected on smaller scales, i.e. when the simulation already fully  resolves the free-streaming scale.} Furthermore, it opens the door to new research areas. Due to phase-space density constraints, the intrinsic velocity dispersion of dark matter is expected to create a core in the innermost part of dark matter profiles \citep{Tremaine:1979}. With this technique, one can, in principle,  resample the velocity dispersion of the dark matter sheet just prior to collapse, and simulate the formation of cores at the centre of the first haloes, getting valuable information such as their exact size and shape.
\footnote{Even if, the primordial velocity dispersion could be added in the initial conditions of standard $N$-body simulations \citep{Colin:2008,Villaescusa-Navarro:2011,Maccio:2012}, it has been shown that this technique introduces significant numerical noise \citep{Leo:2017}.} This may be important for indirect searches for dark matter, since it impacts predictions for dark matter annihilation radiation \citep{Ishiyama:2010,Delos:2022a,Stucker:2023}.

\begin{figure*}
    \centering
    \begin{subfigure}[b]{0.98\textwidth}
   \includegraphics[width=1\textwidth]{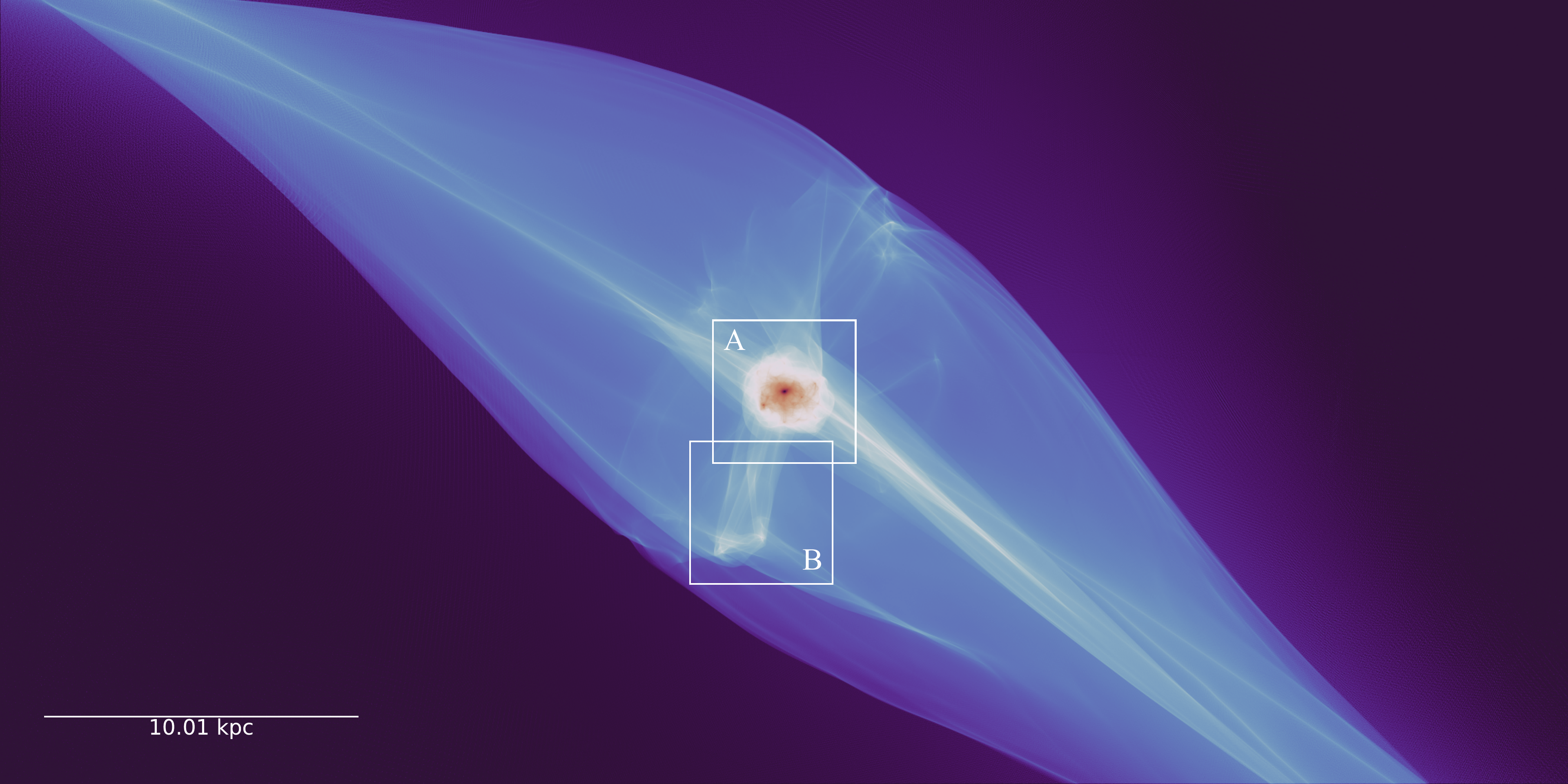}
   %\caption{}
   \label{fig:big_filament} 
\end{subfigure}
\begin{subfigure}[b]{0.49\textwidth}
   \includegraphics[width=\textwidth]{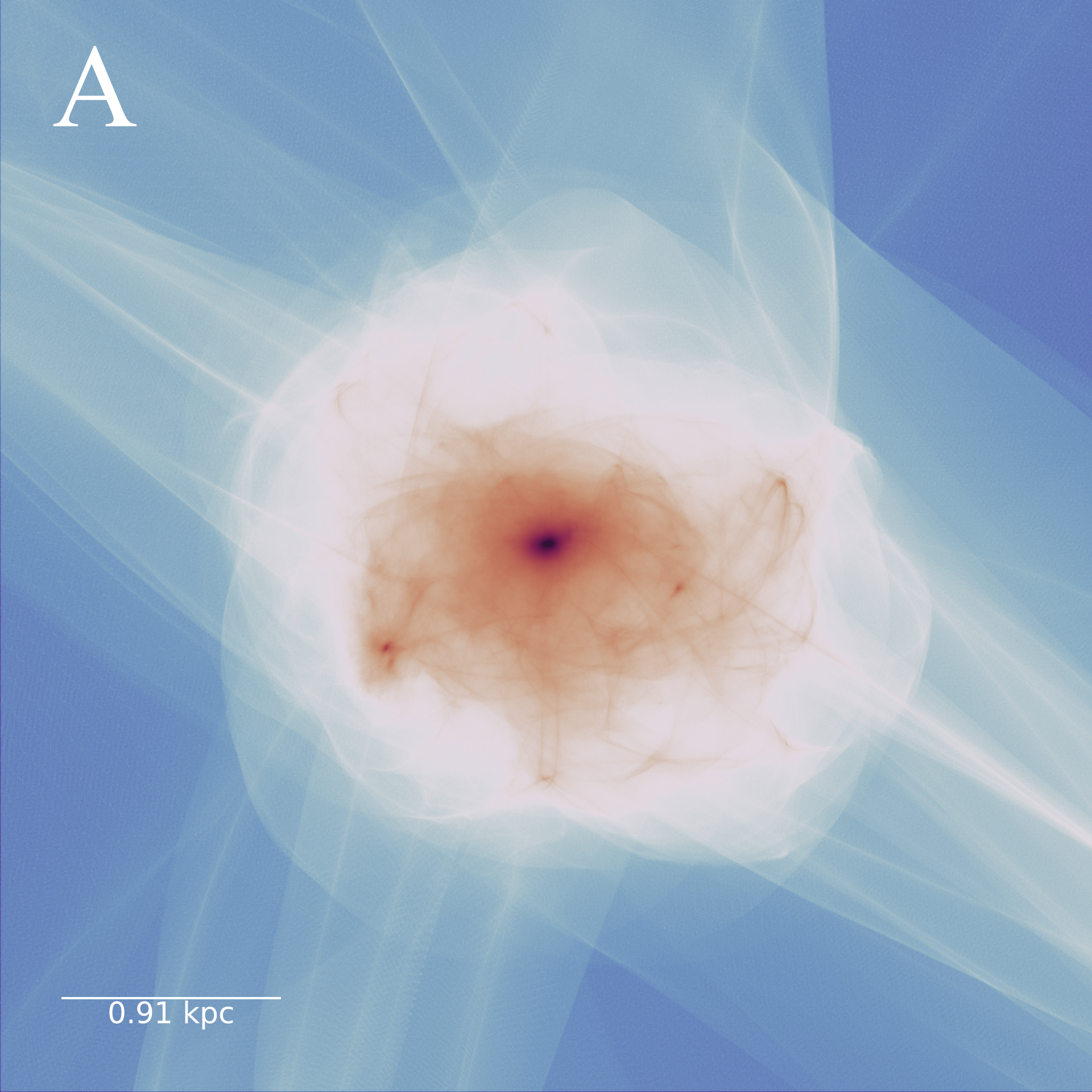}
   %\caption{}
   \label{fig:halo}
\end{subfigure}
\begin{subfigure}[b]{0.49\textwidth}
   \includegraphics[width=\textwidth]{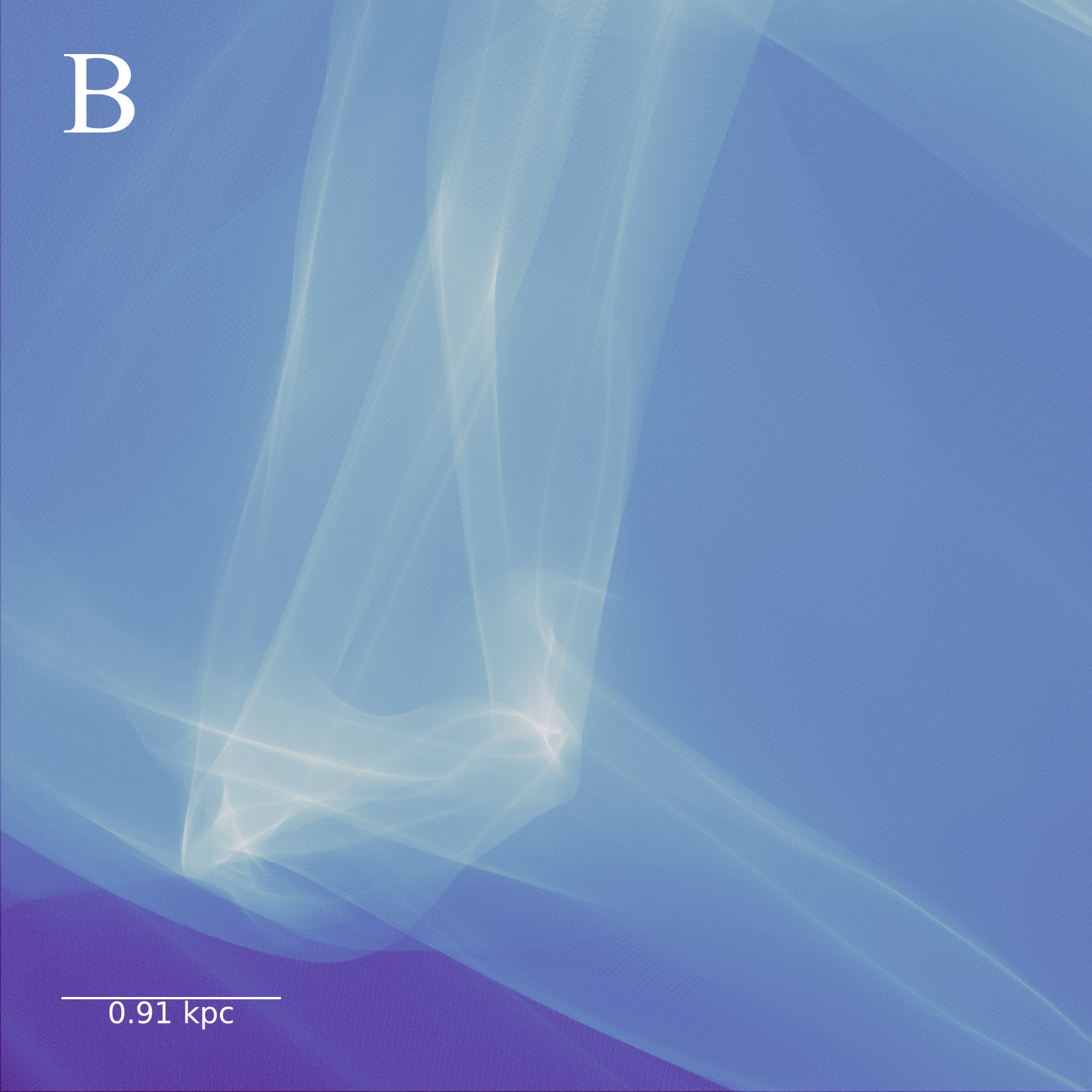}
   %\caption{}
   \label{fig:small_filament}
\end{subfigure}
    \caption{Example of a high resolution sheet-based simulation. The pictures show the formation of a WDM halo in the center of the filament, with no sign of artificial fragmentation. For visualization purposes, we have used the interpolation technique of the sheet presented in \protect\cite{Abel:2012}, even if a significant part of the central halo is already followed with released particles in the simulation. \textit{Upper panel}: a view of the filament that hosts the halo. The projected slice is $63 \rm kpc$ deep, and the picture is $50 \rm kpc$ wide. \textit{Lower left panel}: The forming halo at the center of the filament. \textit{Lower right panel}: Detail of the filament. These last two pictures are $4.5 \rm kpc$ wide and $63 \rm kpc$ deep.}
    \label{fig:sheet_example}
\end{figure*}

In Figure \ref{fig:sheet_example} we show an example of a WDM simulation carried out using this technique at $z=4$. The refinement was done at $z=4.4$. The region that collapsed to the halo was refined three times, increasing the mass resolution by a factor of $8^3$, and reducing the force softening by $2^3$. Once the halo is formed and the sheet becomes too distorted in phase-space, the Lagrangian elements are released and the evolution of the halo is followed with $N$-body techniques. By the time we show the image, $40\%$ of the mass in the halo is in $N$-body particles, while all the rest is deposited in continous Lagrangian elements. Thanks to this, the complex features of the filament are resolved with great detail, without sign of artificial fragmentation. The halo in the center of the filament is resolved with $\sim 2 \cdot 10^7$ particles, with a force softening of $19.53 \rm pc$. The simulation took a total of $\sim 30 $ kCPUh, comparable to a $N$-body simulation at the same mass and force resolution (see Table \ref{tab:sh_nb_num_CPUh}).

\section{Simulations}\label{sec:simulations}

\subsection{Warm dark matter modelling}
\begin{table}
\centering
\begin{tabular}{|c c c c c c|} 
 \hline
    \multicolumn{1}{c}{$m_{\rm DM}$} & \multicolumn{1}{c}{$\alpha$} & $\beta$  & \multicolumn{1}{c}{$k_{\rm hm}$} & \multicolumn{1}{c}{$M_{\rm hm}$} &\multicolumn{1}{c}{$\Delta_{k}$} \\ [0.5ex] 
   [keV]  & $[h^{-1}\rm Mpc]$ & & $ [h \rm Mpc^{-1}]$ & $ [h^{-1} \rm M_{\odot}]$ & $[h \rm Mpc^{-1}]$\\
 \hline
  0.25 & 0.2320 & 2.24 & 1.94& $1.35 \cdot 10^{12}$ & -\\
  1.0 & 0.0498 & 2.24 & 9.05 & $1.33 \cdot 10^{10}$ & - \\
 5.0 & 0.0083 & 2.24 & 54.05 & $6.30 \cdot 10^{7}$ & - \\
 \hline
 1.0 & 0.0185 & 1.0& 9.05 & $1.33 \cdot 10^{10}$ & - \\
 1.0 & 0.0336 & 1.3& 9.05 & $1.33 \cdot 10^{10}$ & - \\
 1.0 & 0.0498 & 1.5& 9.05 & $1.33 \cdot 10^{10}$ & - \\
 1.0 & 0.0707 & 2.24& 9.05 & $1.33 \cdot 10^{10}$ & - \\
 1.0 & 0.0820 & 4.0& 9.05 & $1.33 \cdot 10^{10}$ & - \\
 1.0 & 0.0906 & 9.0& 9.05 & $1.33 \cdot 10^{10}$ & - \\
 \hline
 0.25 & 0.4218 & 9.0 & 1.94 & $1.35 \cdot 10^{12}$ & -\\
 0.25 & 0.4218 & 9.0 & 1.94 & $1.35 \cdot 10^{12}$ & 50 \\
 0.25 & 0.4218 & 9.0 & 1.94 & $1.35 \cdot 10^{12}$ & 25 \\
 0.25 & 0.4218 & 9.0 & 1.94 & $1.35 \cdot 10^{12}$ & 10 \\
\end{tabular}
\caption{The WDM model parameters of our simulations: the dark matter particle mass ($m_{\rm DM}$), scale ($\alpha$) and shape ($\beta$) of the cut-off in the power spectrum, half-mode scale ($k_{\rm hm}$) and half-mode mass ($M_{\rm hm}$). All the simulations have fixed the second shape parameter $\gamma=4.46$. We vary the scale (upper row), as well as the shape (middle row) of the cut-off. Independently (lower row), we keep the cut-off fixed and we add small-scale structure, at scales given by $\Delta_k$.}
\label{table:wdm_models}
\end{table}

In WDM models small-scale perturbations are washed out, and a cut-off is imprinted on the power spectrum. The mass and interaction cross-sections of the dark matter particle determine its production and evolution, setting the scale and shape of this cut-off. Although {\it a priori} each dark matter model imprints its own specific cut-off, \cite{Stucker:2022} showed that the following parametrization of the transfer function \citep{Murgia:2017} adequately captures a vast range of models,
\begin{equation}\label{eq:transfer}
 T(k) := \sqrt{\frac{P_{\chi}(k)}{P_{\rm CDM}(k)}} \simeq (1+(\alpha k)^{\beta})^{-\eta},
\end{equation}
where `$\chi$' stands for a generic WDM candidate, `CDM' labels a perfectly cold model without a cut-off, $\alpha$ is the characteristic length scale of the cut-off, and $\beta$ and $\eta$ control the shape of the cut-off at larger and smaller scales, respectively. If dark matter was produced thermally, $\alpha$ can be related to the mass of the warm dark matter particle ($m_{\rm DM}$) \citep{Viel:2005},
\begin{equation}
 \alpha = 0.049 \left[ \frac{m_{\rm DM}}{\rm keV} \right]^{-1.11} \left[ \frac{\Omega_{\rm DM}}{0.25} \right]^{0.11}\left[ \frac{h}{0.7} \right]^{0.2} h^{-1} \rm Mpc,
\end{equation}
where $\Omega_{\rm DM}$ is the density parameter for dark matter in the universe today, and $h$ is the dimensionless Hubble constant. Finally, it is useful to introduce the \textit{half-mode} wavenumber $k_{\rm hm}$, defined by $T(k_{\rm hm}) = 0.5$. The corresponding mass scale is the half-mode mass $M_{\rm hm}$ \citep{Schneider:2012}
\begin{equation}
 M_{\rm hm} := \frac{4 \pi}{3} \rho_0 \left( \frac{\pi}{k_{\rm hm}} \right)^3,
\end{equation}
where $\rho_0$ is the mean matter density of the universe at $z=0$. This mass-scale is a good estimate of the smallest mass haloes to form. Due to hierarchical structure formation, it is also the mass-scale of the very first haloes. In Table \ref{table:wdm_models} we list the values of these quantities for the models that we will use in this work. 

\subsection{Numerical set-up}

In this work we want to simulate the first and smallest haloes that form in cosmologies with a cut-off in the power spectrum. In order to be able to resolve the inner profile of such haloes in standard cosmological simulations, one has to pick a very small box-size (see, for example, \citealt{Ishiyama:2014}). The advantage of such an approach is that one can do some statistics, given that all the first haloes that form in the box are equally well resolved and can be followed as they grow. However, the effect of perturbations larger than the box are not included and are substantial. Furthermore, the simulation must be stopped as soon as the longest wavelengths that are included approach nonlinearity (for \citealt{Ishiyama:2014}, at $z=32$). Another option is the zoom-in technique, where only a small region of the cosmological box is followed at high resolution. Of course, this option lacks statistics, but it properly includes the effect of long-wavelength perturbations. This is the approach that we have picked for this paper.

The procedure goes as follows: first, we run a $N$-body parent simulation with the minimum resolution needed to resolve the half-mode mass with approximately one thousand particles. The boxsize of these simulations is $L=32 h^{-1}\rm Mpc$, and their resolution is set by the number of particles in a half-mode mass. Then, in a given snapshot at the redshift of interest, we find a halo with $M \sim M_{\rm hm}$, that has just collapsed, and we identify its Lagrangian region. We pay special attention to selecting haloes that have not had any visible major merger, given that mergers could erase the signature of the initial collapse \citep{Angulo:2017, Ogiya:2016}. Finally, we use MUSIC \citep{Hahn:2011} to create initial conditions for the zooms at $z=49$, using 2nd order Lagrangian Perturbation Theory to displace particle positions and velocities from a regular mesh. We solve for the gravitational evolution using two independent schemes. $N$-body simulations are run with Gadget4 \citep{Springel:2021}, and the refined S+R simulation code is implemented on top of a modification of Gadget3 \citep{Stucker:2020}. For both cases, a softening length of $\epsilon = 0.02 l$ is used, where $l$ is the mean interparticle distance in comoving units. 

We have recursively zoomed onto the same Lagrangian regions, increasing the mass resolution by a factor of eight in each iteration. The force resolution is updated accordingly\footnote{It is important to notice that each time we increase  resolution in the zoom-in region, we have to increase it in the rest of the box also. Otherwise, we may get biased results \citep{Onorbe:2014}.}. The numerical parameters of each zoom level are listed in Table \ref{tab:params_zooms}. The names are constructed adding the dark matter particle mass and the zoom level with respect to the resolution of the parent simulation. Following \cite{Onorbe:2014}, we have made sure that our high-resolution region is large enough to be unaffected by low-resolution particles. In Appendix \ref{sec:conv_Nb} we test convergence between simulations at these different resolution levels.

\subsection{Simulation sets}\label{sec:sim_sets}
\begin{table*}
\centering
\begin{tabular}{|c c c c c c c c c | }
\hline
     name & $m_{\rm DM} [\rm keV]$& $z_{\rm col}$&$z_{\rm hm}$  & \multicolumn{1}{c}{set 1} & \multicolumn{1}{c}{set 2} & \multicolumn{2}{c}{set 3} \\%& max zoom level\\
     & & &   & $N$-body+sheet& small scale power & $\beta=2.24$ & vary $\beta$ \\
\hline
0.25kev-zhm0.0 & 0.25& 1.77&0.0 &  - & - & \checkmark & - \\%& 025-L4 \\ 
0.25kev-zhm3.0 & 0.25& 4.2&3.0 & - & \checkmark & \checkmark & -\\%& 025-L4\\ 
0.25kev-zhm2.0 & 0.25& 4.09&2.0& - & - & \checkmark & -\\%& 025-L4\\ 
0.25kev-zhm1.0 & 0.25& 3.67&1.0 & - & - & \checkmark & -\\%& 025-L5\\ 
0.25kev-zhm0.5 & 0.25& 3.23&0.5 & - & - & \checkmark & -\\%& 025-L4\\ 
\hline
1kev-zhm0.0 & 1.0& 2.34&0.0& \checkmark & - & \checkmark & -\\%& 1-L5 \\ 
1kev-zhm3.0 & 1.0& 5.59&3.0 & - & - & \checkmark & \checkmark\\%& 1-L5 \\ 
1kev-zhm2.0 & 1.0& 4.5&2.0 & - & - & \checkmark & -\\%& 1-L5 \\ 
1kev-zhm1.0 & 1.0& 4.82&1.0 & \checkmark & - & \checkmark & -\\%& 1-L6 \\ 
1kev-zhm0.5 & 1.0& 3.37&0.5 & \checkmark & - & \checkmark & \checkmark\\%& 1-L5 \\ 
\hline 
5kev-zhm0.0 & 5.0& 3.43 &0.0 & - & - & \checkmark & -\\%& 5-L4\\ 
5kev-zhm3.0 & 5.0& 7.48&3.0 & - & - & \checkmark & -\\%& 5-L4\\ 
5kev-zhm2.0 & 5.0& 6.16&2.0 & - & - & \checkmark & -\\%& 5-L4\\ 
5kev-zhm1.0 & 5.0& 3.94&1.0 & - & - & \checkmark & -\\%& 5-L4\\ 
5kev-zhm0.5 & 5.0& 2.67&0.5 & - & - & \checkmark & -\\%& 5-L4\\ 
5kev-zhm4.0 & 5.0& 7.12&4.0 & - & - & \checkmark & -\\%& 5-L4\\ 
    \end{tabular}
    \caption{List of haloes that we simulate for this paper. For each halo, we list the dark matter particle mass ($m_{\rm DM}$), the redshift of collapse ($z_{\rm col}$) and the redshift when the halo reaches the half-mode mass ($z_{\rm hm}$). In the first set, we ran simulations from three initial conditions using both sheet-based and $N$-body techniques (see Table \ref{tab:sh_nb_num}). In the second set, we re-simulate the formation of one particular halo, adding small scale noise to investigate the effect of fragments in a controlled manner. The third set consists of $N$-body simulations of all the haloes, varying the dark matter particle mass and the shape of the cut-off.}
   \label{tab:halos}
\end{table*}

For this paper we have simulated the formation of a total of 16 haloes. We have divided these simulations into three sets. In Table \ref{tab:halos} we specify the name of each halo, the dark matter particle mass ($m_{\rm DM}$), the collapse and half-mode-mass redshifts ($z_{\rm col}$ and $z_{\rm hm}$), and the simulation sets in which each halo has been simulated. The names of the haloes are given by the dark matter particle mass and the redshift at which they reach the half-mode mass. 

The purpose of the first set is to test the role of artificial fragmentation in the formation of the first haloes. It consists of three haloes in the $m_{\rm DM}=1 \rm keV$ cosmology, simulated with both $N$-body and sheet-based techniques. We match the initial conditions, mass resolution and softening length in the two simulation types. We stop these simulations once the halo has reached $\sim 10-20\%$ of the half-mode mass. The numerical parameters are listed in Table \ref{tab:sh_nb_num}. Remarkably, the computational cost of the sheet-based and $N$-body simulations is comparable, of the order of 1k-10k CPUh (see Table \ref{tab:sh_nb_num_CPUh}), depending on the highest resolution we achieve. 

The second set consists of $N$-body simulations of the formation of one particular halo with added amounts of small-scale power in the initial conditions. The halo lives in a cosmology with $m_{\rm DM}=250\rm eV$ and reaches the half-mode mass at $z_{\rm hm} = 3.0$. These simulations are at our fiducial resolution, 025-L4, with particle mass $m_{\rm p} = 3 \times 10^5 h^{-1}\rm M_{\odot}$ (resolving the half-mode mass with $\sim 4 \cdot 10^6$ particles), and  softening length  $\epsilon = 0.312 h^{-1} \rm kpc$ in comoving units. Here also we have done some convergence tests, which are presented in Appendix \ref{sec:conv_noise}. 

Finally, we ran a large set of standard $N$-body simulations to follow the formation of all the haloes listed in Table \ref{tab:halos}. In particular, we vary the dark matter particle mass ($m_{\rm DM}$), the redshift when the halo reaches the half-mode mass ($z_{\rm hm}$), and the  shape of the cut-off ($\beta$). We follow the evolution of the haloes until they have acquired the half-mode mass, $M_{\rm 200c}=M_{\rm hm}$. The fiducial resolution levels for this set of haloes are 025-L4, 5-L4 and 1-L5 in the $m_{\rm DM} = 0.25, 5.0$ and $1\rm keV$ cases, respectively. In the first two, we resolve the haloes at the half-mode mass with $\sim 5 \cdot 10^6$ particles, while in the last one we reach $\sim 4 \cdot 10^7$ particles.

\section{Haloes without artificial fragmentation}\label{sec:halos}
\begin{table*}
    \centering
    \begin{tabular}{|c ccc ccc ccc c c c c| }
\hline
         & \multicolumn{3}{c}{$m_p [\rm M_{\odot}]$} & \multicolumn{3}{c}{$\epsilon (\rm pc)$} & \multicolumn{3}{c}{$N_p[10^6]$} & $z_i$ & $z_{\rm ref}$ &$z_f$ & $M_{f} / M_{\rm hm}$ \\
         & base & refine & $N$-body & base & refine & $N$-body & base & refine & $N$-body & & & & \\
\hline 
1kev-zhm1.0 & $3.65 \cdot 10^4$ & $71.33$ & $71.33$ &  312.5 & 19.53 & 19.53 & $0.76$ & $68.7$ & $277$ & 49.& 4.55 & 4.0 & 0.11\\
1kev-zhm0.5 & $3.65 \cdot 10^4$ & $570.68$ & $570.68$ &  312.5 & 39.06 & 39.06 & $16.7$  & $27.30$ & $79.5$ & 49. & 3.55 & 2.7 & 0.21\\
1kev-zhm0.0 & $4.56 \cdot 10^3$ & $570.68$ & $570.68$ & 156.25& 39.06 & 39.06 & $12.27$ & $11.85$ & $89.03$ & 49.&  2.44 & 2.03 & 0.13\\
    \end{tabular}
    \caption{Numerical parameters of the simulations of set 1: high resolution particle mass ($m_p$), softening length in comoving units ($\epsilon$), number of high resolution particles ($N_p$), initial ($z_i$), refinement ($z_{\rm ref}$) and final ($z_{f}$) redshifts, and the mass of the halo at $z_f$ in units of the half-mode mass ($M_f / M_{\rm hm}$). We point the values for the base and refined sheet simulations, as well as the equivalent $N$-body simulation. The initial conditions are the same for the sheet and $N$-body simulations.}
    \label{tab:sh_nb_num}
\end{table*}

In this section we present simulations of first halo formation with no artificial fragmentation (see also \cite{Colombi:2021}). These very high resolution simulations are enabled by the new numerical techniques introduced in previous sections. Comparing them to standard $N$-body simulations of the formation of the same objects allows us to discern whether artificial fragments significantly affect the internal structure of the first haloes. 

The set consists of three haloes in the $m_{\rm DM} = 1 \rm keV$ cosmology that first collapse at different redshifts (check Table \ref{tab:halos}, set 1). In all cases, a filament forms first, and the halo eventually collapses near its centre. We stop the simulations when the haloes have acquired $\sim 10-20 \%$ of the half-mode mass; at this time their inner structure is set and stabilized. 

\begin{figure*}
\centering
\includegraphics[width=1.\textwidth]{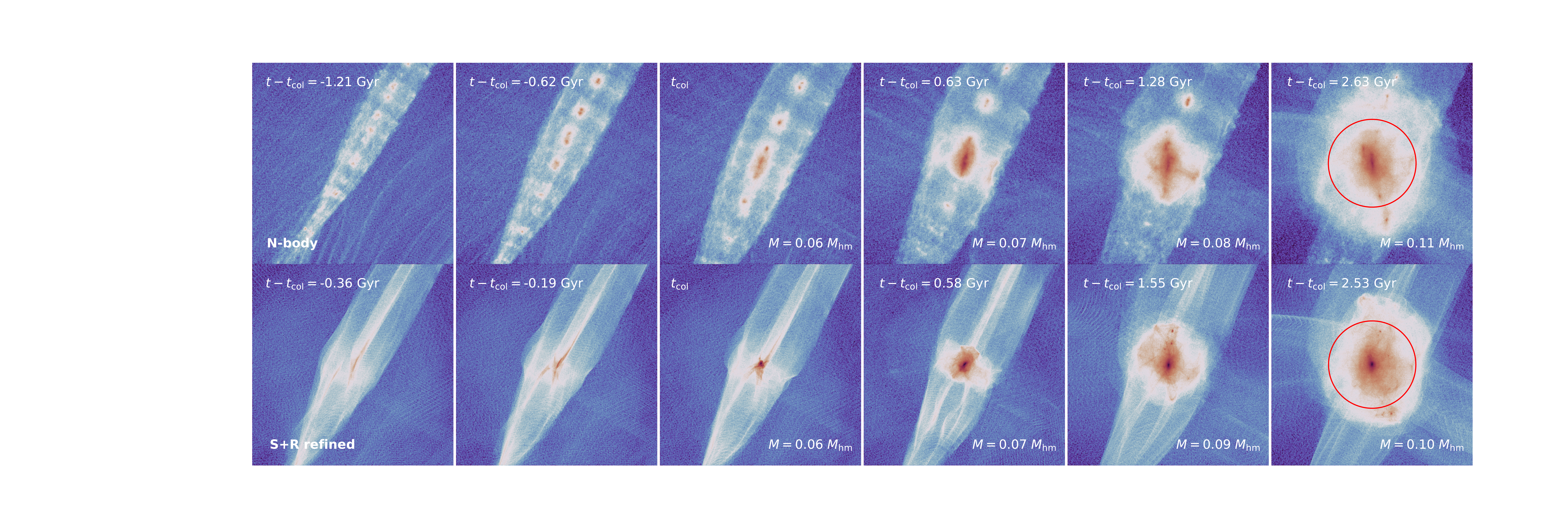}
\caption{The collapse of the halo ``1kev-zhm1.0'' in the $N$-body (upper row) and sheet-based (lower row) simulations. The region we show is $22.7 \rm kpc$ wide in comoving units. The image is made by direct projection of the particles without any interpolation. By the time we stop the simulation at $z \sim 4.03$, the halo has evolved for $\sim2.58 \rm Gyr$, and it has acquired $10\%$ of the half-mode mass. The red circle denotes the $r_{\rm 200c} / 8$ of the halo at this moment. For comparison, the same region is indicated in Figure \ref{fig:shvsnbdensity}.}
    \label{fig:shvsnbcollapse_5}
\end{figure*}

\begin{table}
    \centering
    \begin{tabular}{|c ccc | }
\hline
         &  \multicolumn{3}{c}{CPUh}  \\
         & base & refine & $N$-body  \\
\hline 
1kev-zhm1.0 & $3.45 \cdot 10^2$ & $3.36 \cdot 10^4$ & $3.12 \cdot 10^4$ \\
1kev-zhm0.5 & $1.09 \cdot 10^3$ & $1.1 \cdot 10^4$ & $3.89 \cdot 10^3$ \\
1kev:zhm0.0 & $3.99\cdot 10^3$ & $6.35 \cdot 10^3$ & $4.03 \cdot 10^3$  \\
    \end{tabular}
    \caption{Comparison of the computing cost of the sheet based and $N$-body simulations. The total time of the sheet-based simulations is the sum of the base and refine. The values correspond to the runs of three haloes presented in Section \ref{sec:halos}. In Table \ref{tab:sh_nb_num} we list the numerical details of the simulations.}
    \label{tab:sh_nb_num_CPUh}
\end{table}
\subsection{Artificial fragmentation}

In Figure \ref{fig:shvsnbcollapse_5} we show different snapshots of the formation of the halo ``1kev-zhm1.0''. Even though we do not expect any physical small-scale structure in the filaments before collapse of the halo, we find that the filament in our $N$-body simulation (upper row) contains many small dense clumps by then -- artificial fragments. The fragments appear more than a Gyr before fully three-dimensional collapse of the initial peak. They are evenly spaced along the filament and their mass depends on resolution: at higher resolution, the fragments are smaller, but there are more of them. By the time the halo collapses, the fragments have merged together to become much more massive. Thus, at the times we are interested -- during halo collapse -- , the resolution dependence of fragment mass is not evident \citep[see][]{Wang:2007}. % https://www.youtube.com/watch?v=C9QBbAIAGb8&ab_channel=LurdesOndaroMallea    https://www.youtube.com/watch?v=8ChIUCE4NAw&ab_channel=LurdesOndaroMallea

The presence of artificial fragments may significantly alter the halo formation process. The collapse looks very different in the two cases of Figure \ref{fig:shvsnbcollapse_5}. The halo forms almost exclusively through mergers of fragments in the upper row. In the sheet-based simulation, however (the lower row) the filament is completely smooth, and the halo collapses monolithically at at the filament's centre. In these first stages, around $\sim 0.6 \rm Gyrs$ after the collapse, the haloes look different in the two simulations. Unlike the halo in the sheet-based simulation, which is nearly spherical and has a well-defined center, the $N$-body halo is very elongated in the filament direction, retaining the memory of the mergers. As the haloes grow, these differences seem to decrease, and the haloes become more similar. 

\begin{figure*}
 \centering
\includegraphics[width=\textwidth]{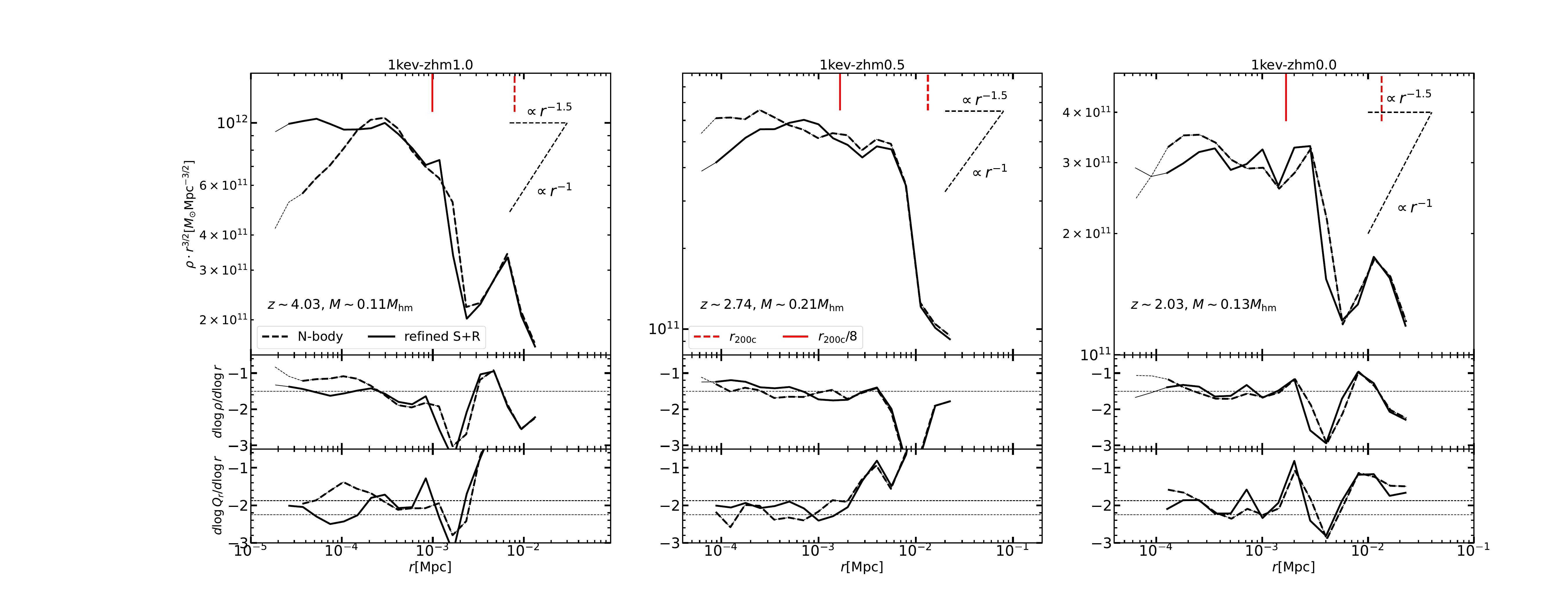}
    \caption{The density profiles measured in $N$-body (dashed) and sheet-based (solid) simulations. In the upper row we display the density profile multiplied by $r^{1.5}$, and in the middle row the logarithmic slope of the density profile. The lower row shows the logarithmic slope of the pseudo phase-space density $Q_r$. Each panel corresponds to a halo. In the first halo -``1kev-zhm1.0''- the pre-collapse filament is strongly fragmented in the $N$-body simulation. The filament of the second halo -``1kev-zhm0.5''- is very complex, and physical mergers happen in the collapse. In the last halo -``1kev-zhm0.0''- the filament in the $N$-body simulation does not fragment, and no physical mergers happen. We show the profiles down to the resolution limit of our simulations, given by the maximum of the convergence radius \protect\citep{Power:2003} and three times the softening length. In the thin lines we correct the convergence radius, taking into account the recent formation of the halo. The red vertical lines denote the $r_{\rm 200c}$ (dashed) and $r_{\rm 200c}/8$ (solid) radii. The latter serves as reference to the red circle in Figure \ref{fig:shvsnbcollapse_5}, for the halo 1kev-zhm1.0.}
    \label{fig:shvsnbdensity}
\end{figure*}

In the left panel of Figure \ref{fig:shvsnbdensity} we display the density profile of this halo in the last snapshot, as measured in the $N$-body simulation (dashed lines) and in the sheet-based simulation (solid line). The profiles are shown down to the numerical resolution of our simulations, given by the maximum of the softening radius ($r_{\rm soft} = 3 \epsilon$) and the convergence radius \citep{Power:2003}. In the thin lines we have corrected the convergence radius, taking into account the recent formation of the halo. The standard estimation from \cite{Power:2003} computes the 2-body relaxation within a halo in the age of the Universe. As these haloes have formed very recently, the particles have had much less time to relax due to 2-body interactions. At first order, the correction factor goes like $(t_{\rm a} - t_{\rm col})/t_{\rm a}$, where $t_{\rm col}$ and $t_{\rm a}$ are the age of the Universe at the time of collapse and at the moment where we compute the profiles, respectively. Unless stated the contrary, we use the same cuts in all the density profiles that we show in this work. 

Comparing the $N$-body and sheet-based outputs, we distinguish two regimes. For $r>2~\rm kpc$, the two profiles agree. For $r<2~\rm kpc$, though, the halo in the sheet-based simulation forms a steep cusp $\propto -1.5$, while in the $N$-body simulation it has a much shallower profile. This is a clear signature of artificial fragmentation. For the first time we can reliably state that it changes the formation and structure of the haloes in a non-negligible way. Indeed, together with the 2-body relaxation and force softening, this is a new mechanism that modifies the inner structure of haloes in $N$-body simulations for numerical reasons.

The effect of artificial fragmentation is similar to that of numerical resolution: it shallows the inner density profiles. However, the mechanisms at play are presumably very different. As is visible in Figure \ref{fig:shvsnbcollapse_5}, the halo forms when the artificial fragments merge together. That is: the initial phase of  collapse involves major mergers of artificial fragments. Interestingly, several earlier papers have found that the violent relaxation during repeated major halo mergers drives the slope of the profile towards shallower values \citep{Ogiya:2016,Angulo:2017}. We expect the same process to take place when artificial fragments, rather than standard haloes, merge. Hence, we speculate that the numerical shallowing of the inner density profiles in $N$-body simulations with artificial fragments can be caused by violent relaxation induced by their mergers. Another possible explanation for this effect could be that the fragmentation leads to dilution of the coarse-grained phase-space density which then cannot be compressed as effectively as in the unfragmented case. It is plausible that in this case a phase space diffusion process driven by scattering off of both fragments is at work. We will explore these ideas further in Section \ref{sec:artfrags} and Appendix \ref{sec:phase_space}.

Moreover, we have also looked at other halo properties, such as velocity dispersion profiles, velocity anisotropy, shapes, and profiles of the pseudo phase-space density. Our findings are consistent with previous work (see, for instance, \cite{Delos:2022}). In particular, no clear difference is found between the $N$-body and sheet-collapse cases for the velocity anisotropy of prompt cusps (they are approximately isotropic) nor for their shapes, which have similar axial ratios and orientations to the larger scale haloes which form around them.

In the lower panels of Figure \ref{fig:shvsnbdensity}, we show the logarithmic slope of the pseudo phase-space density profiles of the first haloes, $Q \propto \rho / \sigma^3$. In general, in absence of artificial fragmentation, they follow a power-law $Q \propto r^{\chi}$, with $\chi < -1.875$, reaching to $\chi \sim -2.25$ in some cases. Interestingly, for the $N$-body simulation of 1kev-zhm1.0, where the fragmentation is severe and the halo has formed though mergers of fragments rather than monolithic collapse, $\chi \sim -1.875$. Yet, these plots are not conclusive. We will explore this further in Section \ref{sec:artfrags_Nb}.

\subsection{Poorly resolved filament}

\begin{figure}
\centering
\includegraphics[width=0.8\columnwidth]{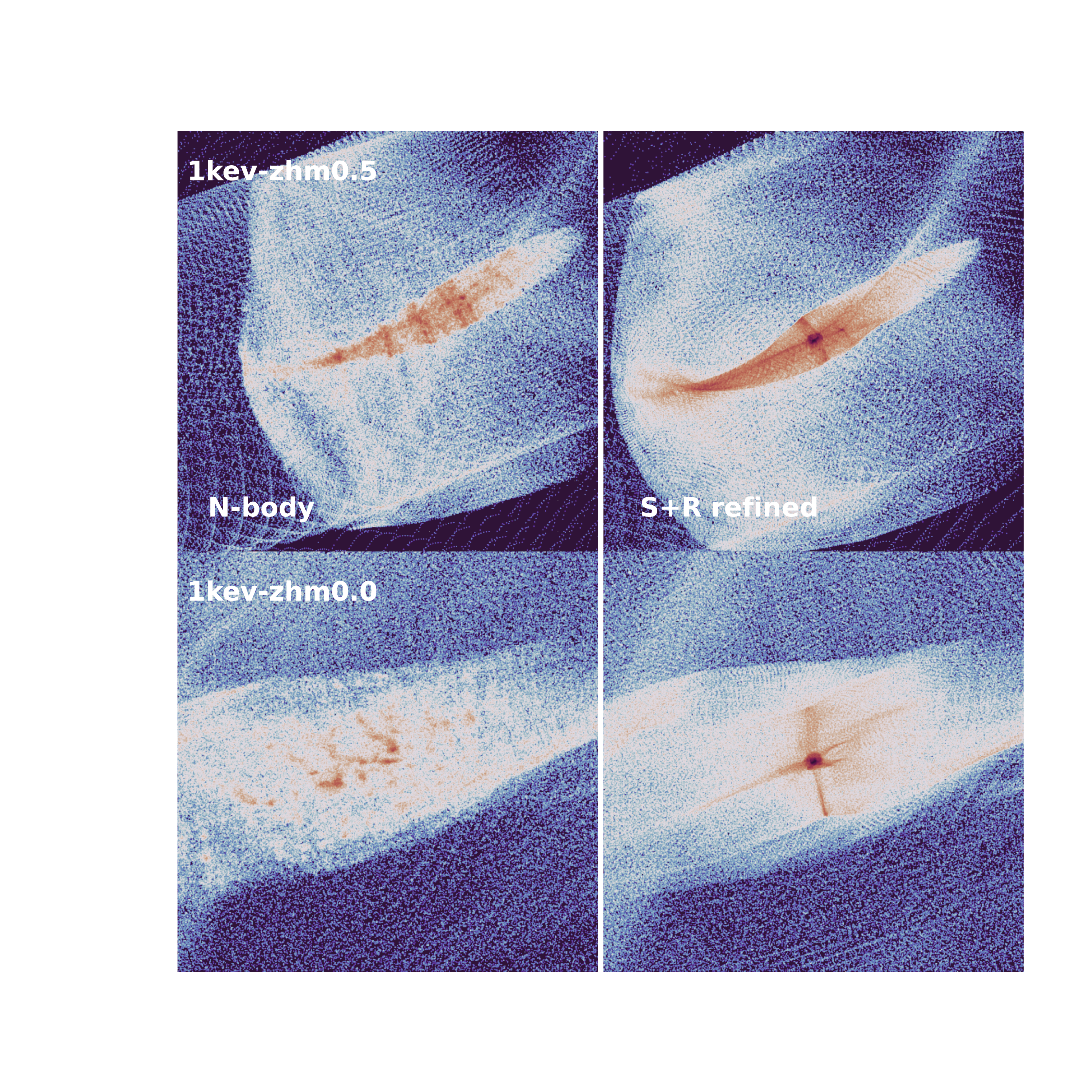}
\caption{Density field of the filament at halo formation, ``1kev-zhm0.5'' at $z=3.5$ (upper row) and ``1kev-zhm0.0'' at $z=2.3$ (lower row). We show the outputs of the $N$-body (left column) and sheet-based (right column) simulations. The region is $31.8 \rm kpc$ wide in comoving units. The image is done by projecting the particles, without any interpolation for visualization purposes.}
    \label{fig:precoll_poordens}
\end{figure}

The discretization of the dark matter sheet in $N$-body simulations yields a noisy density estimate, especially in low-density regions -- i.e where all the mass of a large region is deposited in a few $N$-body particles. In the sheet-based approach, the mass is distributed rather than concentrated in point particles, leading to a much better density estimate. Filaments are a good example of this, as is clear from Figure \ref{fig:shvsnbcollapse_5}. For instance, at $t-t_{\rm col} \sim 0.6 \rm Gyr$, the sheet-based simulation shows a filament with a pronounced inner caustic, following the axis of the filament on one side of the halo, and divided in two in the other side. All these features are missing in the $N$-body simulation of the same filament. In their place, the axis of the filament is plagued with fragments. 

However, not all filaments fragment in $N$-body simulations. In Figure \ref{fig:precoll_poordens} we show the density field of the filament at its collapse time for two further haloes of our set, ``1kev-zhm0.5'' and ``1kev-zhm0.0'', as resolved in $N$-body (left) and sheet-based (right) simulations. It is clear that the $N$-body density estimate is very noisy in both filaments. Arguably, they are in the process of fragmenting. Yet, the halo forms before this process is finished, and prevents the formation of virialized clumps -- fragments. In general, the factors that drive or prevent the formation of fragments are not clear, but we speculate that they have to do with the tidal field exerted in the filament. The stronger the tidal field, the more material will be flowing in the uncollapsed direction, and it will be harder to create fragments. In contrast, if the filament is very calm and quiet, it will be easier for the discreteness errors to grow and collapse into fragments. Similarly, we expect the filaments that have formed much before the formation of the first halo in them -- as ``1kev-zhm1.0''--, to be more likely to fragment than those where the halo forms soon after the filament -- as in ``1kev-zhm0.5'' and ``1kev-zhm0.0''. 

Even in the absence of fragments, the poorly resolved filament alters the small scales of halo formation, leading to slightly different scenarios in $N$-body and sheet-based simulations (Figure \ref{fig:precoll_poordens}). In the former, the sharp features and edges of the filaments are blurred, and so is the initial collapse of the haloes. At the same mass and force resolution, the sheet-based simulations display a well-defined condensed cusp forming in the center of the smooth filament. These differences can extend after the initial collapse. The morphology of the filament of ``1kev-zhm0.5'' is quite complex, it is folded, and a small physical clump forms in the intersections. The $N$-body simulation is not able to capture these sharp features of the filament, and the resulting clump is fluffier than in the sheet-based simulation. Soon after having formed, this clump merges with the central halo. 

In the central and right panels of Figure \ref{fig:shvsnbdensity} we show the density profiles of the haloes ``1kev-zhm0.5'' and ``1kev-zhm0.0'', respectively. Despite the differences we have mentioned, the structure of the resulting haloes agrees relatively well in the sheet-based and $N$-body simulations: in both cases, the haloes form a steep cusp $\propto -1.5$. In the innermost region, ``1kev-zhm0.5'' seems to be shallower in the sheet-based simulation than in the $N$-body simulation. The discrepancy between the two simulation techniques is not as strong as in ``1kev-zhm1.0'', where the pre-collapse fragmentation is very severe. We argue that the merger with the aforementioned second clump could be responsible for this: the clump is resolved better in the sheet-based simulation, thus, the merger with the central halo is stronger, and the inner cusp is more efficiently shallowed \citep{Angulo:2017,Ogiya:2016}. However, this is just a guess: a larger set of simulations is needed to confirm these hypotheses.

In summary, the formation of prompt cusps in the central regions of the first haloes is firmly established by our simulations. Independent simulation schemes, sheet-based and standard $N$-body simulations, point to this result. 
Nevertheless, differences in the small scales of the pre-collapse field can alter the collapse and innermost structure of the halo noticeably. In particular, when the dark matter filament spuriously fragments before the collapse of the halo that will form in it, the process can be greatly affected. Importantly, we see a decreased slope of the inner density profile due to the presence of fragments in $N$-body simulations. In the following section we explore this issue further. 

\section{Understanding the role of artificial fragmentation}\label{sec:artfrags}

Artificial fragments arise due to the discrete sampling of the dark matter sheet. While a thorough study of their formation and structure is outside the scope of this work, we next investigate how they influence the collapse of the first haloes. For this purpose, we think of them as small-scale perturbations (effectively noise) in the smooth dark matter field, irrespective of details of their origin.

In this section we introduce a second set of simulations designed to explore this issue. These are $N$-body resimulations of the formation of a single halo with different amounts of added small-scale power. We adjust the degree of small-scale nonlinear structure prior to halo collapse in order to explore its effect on halo profiles. In addition to gaining insight into the physics at play, this allows us to constrain empirically the region of the profiles that is modified by such small-scale structure.

\subsection{Power spectra}

\begin{figure}
    \centering
    \includegraphics[width=0.8 \columnwidth]{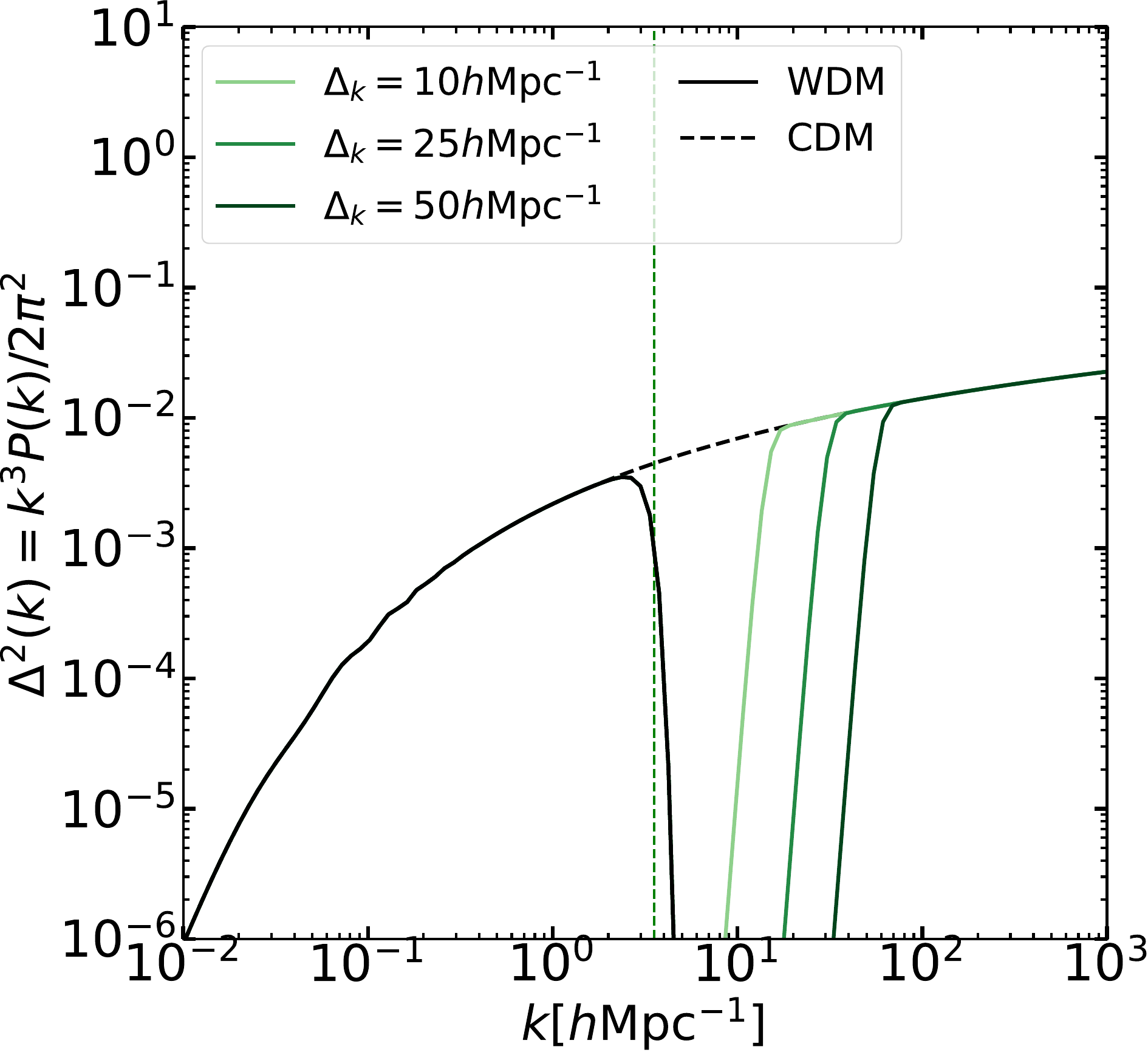}
    \caption{Dimensionless power spectra of  an $m_{\rm DM} = 250 \rm eV$ and $\beta=9$ cosmology modified to include different amounts of small-scale power. The gap $\Delta_k$ sets the distance from the WDM cut-off to the onset of added small-scale power which follows a CDM spectrum. The vertical dashed line represents the half-mode scale.}
    \label{fig:powerspec_noise}
\end{figure}

The initial (linear) power spectrum of our controlled simulations is a standard CDM spectrum, where we zero the power at intermediate scales. By varying the extent of the power 'gap', we control the scale and amplitude of the additional power. The transfer function is given by  expression:
\begin{equation}\label{eq:noise_transfer}
   T_{\rm sm}(k) := T_{\rm WDM}(k) + T_{\rm noise}(k) 
\end{equation}
\noindent where $T_{\rm WDM}(k)$ and $T_{\rm noise}(k)$ follow Equation \ref{eq:transfer},
\begin{equation}\label{eq:noise_transfer2}
   T_{\rm sm}(k, \alpha_1, g_{\rm amp}; \alpha, \beta, \eta) := T(k;\alpha, \beta, \eta) + g_{\rm amp}(1-T(k; \alpha_1, \beta, \eta)).
\end{equation}
The first term imposes a WDM cut-off, while the second introduces additional small-scale power. the amplitude of the added power is regulated by $g_{\rm amp}$ and its scale by $\alpha_1$. Defining $\Delta_k$ as the $k$ distance between the cut-off and the small-scale power, we can obtain $\alpha_1$ via,
\begin{equation}
    \alpha_1(\Delta_k, \alpha, \beta, \eta) = \left(\frac{1}{\alpha} + \frac{\Delta_k}{(2^{1/\eta}-1)^{1/\beta}} \right)^{-1}.
\end{equation}

In Figure \ref{fig:powerspec_noise} we show power spectra for three values of $\Delta_k$, at fixed $g_{\rm amp}=1$, the only value we consider in this paper. The corresponding values are listed in Table \ref{table:wdm_models}. As we increase $\Delta_k$, the power is at smaller scales. If $\Delta_k$ is arbitrarily large, we get a WDM spectrum (black solid line). Similarly, if $\Delta_k=0$, we recover a pure CDM scenario (black dashed line).

This type of experiment can be useful for understanding the transition between the NFW-like structures that emerge in CDM, and the power-law cusps that we find forming promptly in first haloes. There are strong reasons to believe that mass accretion histories play an important role: self-similar assembly of mass in normal halo formation, and monolithic collapse in prompt cusp formation. Nevertheless, it is interesting to study the evolution of the profile as we decrease $\Delta_k$ at fixed $g_{\rm amp}$, or decrease $g_{\rm amp}$ at fixed $\Delta_k$, in a similar way as \cite{Ogiya:2016}. Understanding the transition between the two regimes would shed light on their physical origin, and improve our comprehension of structure formation. Here, we adopt this set-up to explore the imprint that small-scale structure -- such as artificial fragmentation-- leaves in the inner structure of the first haloes. 

\subsection{Haloes with small scale noise}\label{sec:small-scale-noise}

\begin{figure}
    \centering
    \includegraphics[width=0.8\columnwidth]{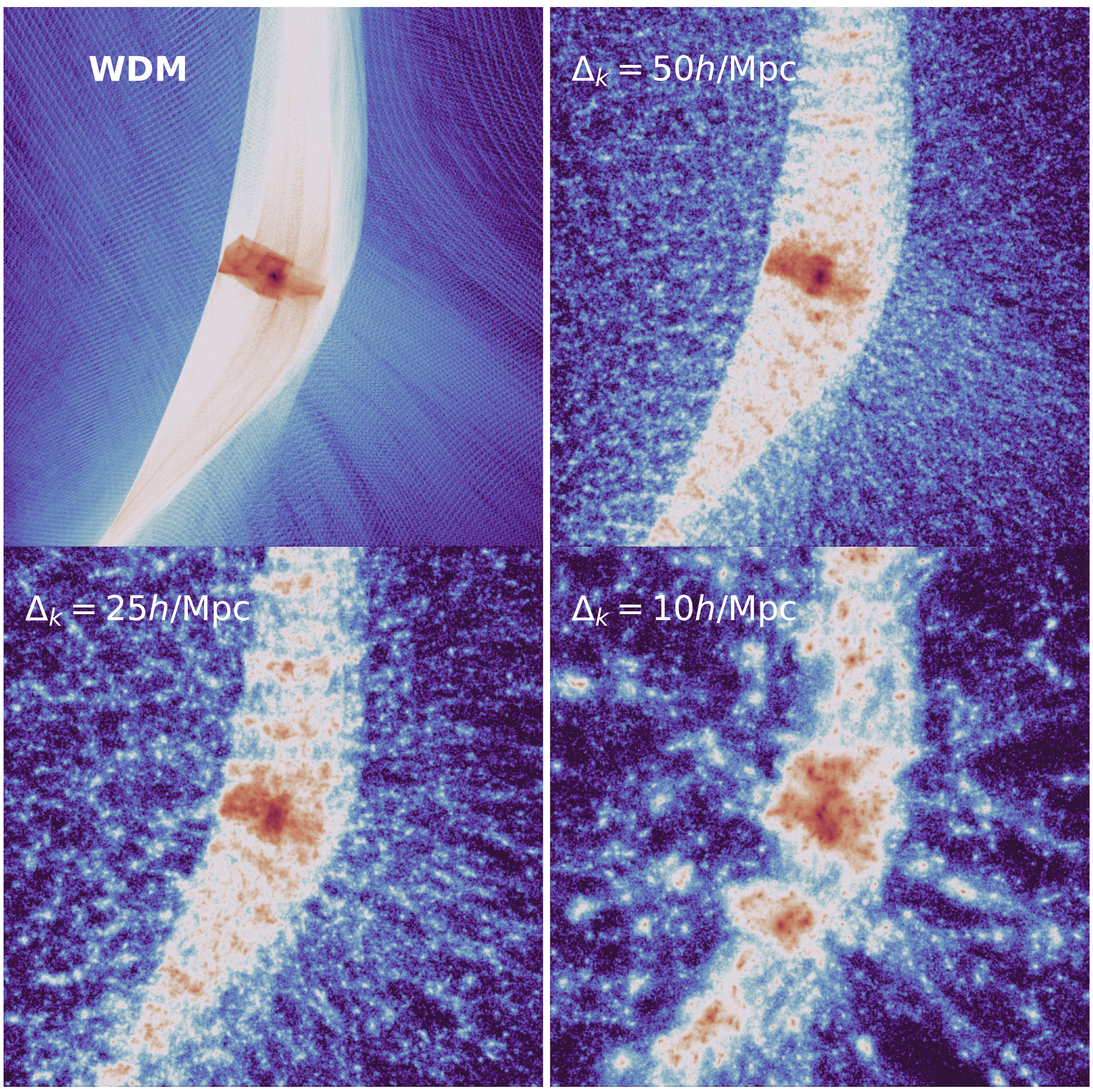}
    \caption{The same halo collapsing at $z=4.56$, in a pure warm cosmology (upper left), and with added small-scale power (upper right and lower panels). We project a slice of $0.1~\rm Mpc$ thickness. The width of the panels is $0.245~\rm Mpc$ in physical units. The halo in the upper left has not fragmented, or the fragmentation is very weak. In the other cases, the degree of small scale structure is controlled by $\Delta_k$.}
    \label{fig:noise_warm2_obj3}
\end{figure}

\begin{figure}
    \centering
    \includegraphics[width=0.9 \columnwidth]{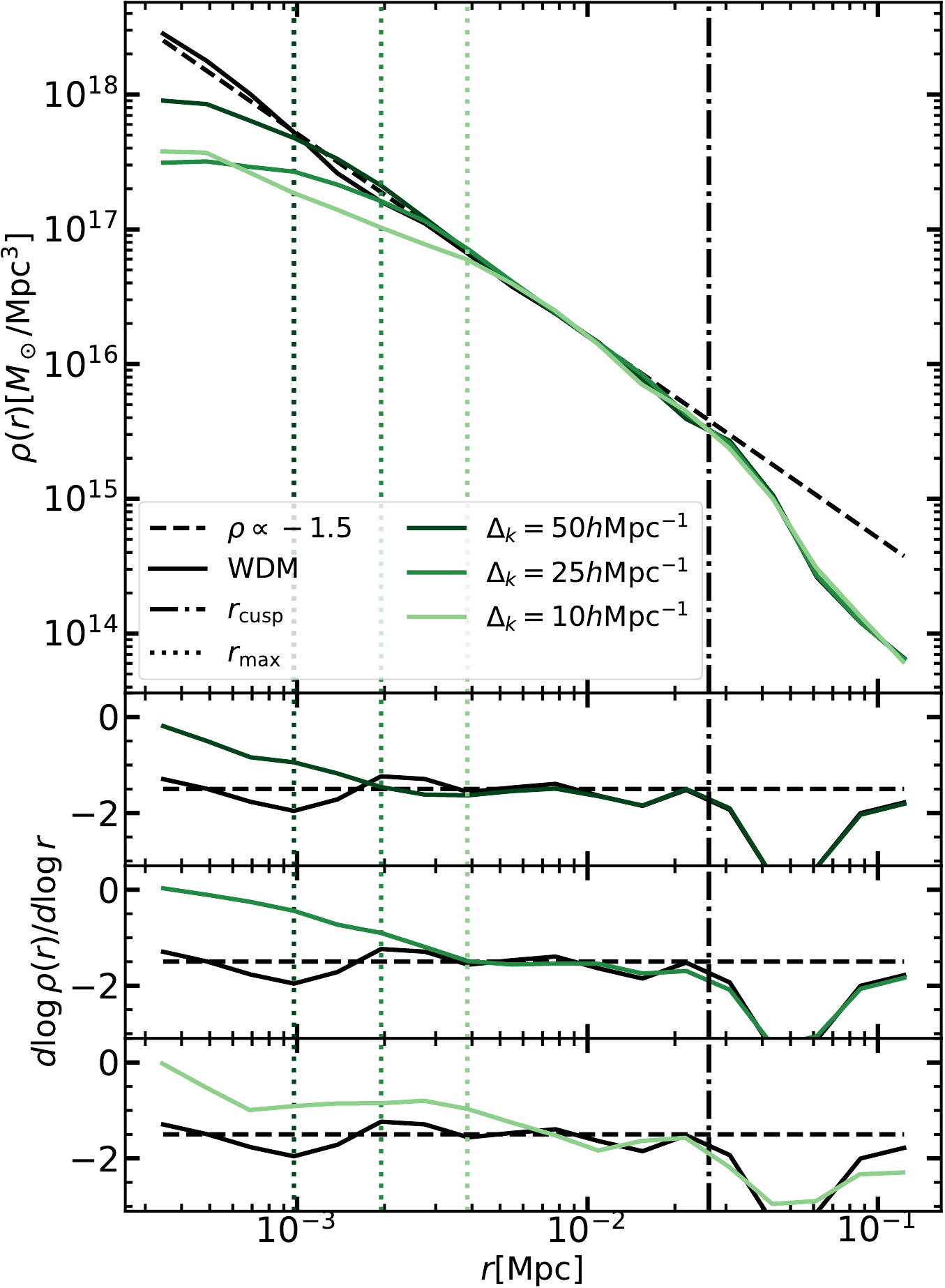}
    \caption{Density profiles of the haloes shown in Figure~\ref{fig:noise_warm2_obj3} but at $z=4.0$. The pure WDM case is a black solid line, while simulations with added small-scale power give  the green solid lines. The upper panel shows the density in physical units, the three lower panels the logarithmic slope of these profiles. The dashed black line indicates $\rho \propto -1.5$, and the vertical dashed-dotted line, the cusp radius given by \protect\cite{Delos:2019}. The dotted lines denote $r_{\rm max}$, the radii where the profiles become $20\%$ shallower than the $\rho \propto -1.5$ power-law.}
    \label{fig:noise_dens}
\end{figure}

The second simulation set consists of four versions of the collapse of the same halo: for a pure WDM spectrum; adding small-scale power with $\Delta_k=50, 25$ and $10 h\rm Mpc^{-1}$. The numerical details of the simulations are described in Section \ref{sec:sim_sets}. In Figure \ref{fig:noise_warm2_obj3} we display the dark matter field at the moment of collapse. It is clear that the haloes collapse at the same time in all four, and that the collapsing region has the same overall structure. In other words, the small-scale structure has not altered the overall mass accretion history of the halo and the collapse remains monolithic. However, the field from which the halo collapses is full of small clumps when power has been added. Depending on the value of $\Delta_k$, the clumps are smaller (top right panel) or larger (bottom right panel). Our hypothesis is that these small clumps influence the collapse of the halo in a similar way to artificial fragments in $N$-body simulations of first halo formation. 

With this we do not mean that these clumps are like artificial fragments. Indeed, they are standard NFW haloes of a CDM cosmology that are found in all the volume, have a spectrum of masses, and are irregularly spaced; whereas artificial fragments form in a completely different way, appear only in the filaments, evenly spaced, all have a similar mass and presumably do not follow a NFW profile. Yet, they are both virialized objects that alter the small scales of the collapse of a halo, without changing the large scale features, such as the mass accretion rate. Arguably, the specifics of their inner structure and distribution in space are not relevant, at least at first order. 

The spherically averaged structure of the resulting haloes is depicted in Figure \ref{fig:noise_dens}, the density profiles as a function of radius in the upper panel, and their logarithmic slope in the three lower panels. As expected, in the absence of small-scale power, the quasi-equilibrium part of the halo follows a power law,  $\rho \propto r^{-1.5}$ (solid black line compared to the straight dashed line). The haloes that have formed from a clumpy field (the green solid lines) agree with this profile at larger radii, but in the inner regions they become shallower, deviating from the power-law form. Qualitatively, this is the same behaviour that we observe in $N$-body simulations of first halo formation when artificial fragments are significant during the collapse.

It therefore seems that the presence of small pre-collapse clumps does not modify the profile at all radii, but only in the inner regions. Interestingly, the maximum scale that is affected correlates with $\Delta_k$: larger pre-collapse clumps affect the density profile out to larger radii. This scale seems to be stable: it does not change with redshift nor with the resolution of our simulations (c.f. Appendix \ref{sec:conv_noise}). 

In the modified region, the behaviour of the profiles varies with $\Delta_k$. As is visible in the last row of Figure \ref{fig:noise_dens}, the density profile of the halo with the largest clumps approaches $d \log \rho / d \log r \sim -1$ in the smallest scales. In contrast, when the clumps are smaller, the haloes appear to form a core (second and third rows of Figure \ref{fig:noise_dens}). It is not clear what causes these differences. A reason could be the resolution of our simulations: the clumps are not equally well resolved in all cases. The largest clumps have $\sim 13000$ particles in $\Delta_k=10h/\rm Mpc$, while in $\Delta_k=50h/\rm Mpc$ they are resolved with $\sim 500$ particles. We test the effect of resolution in Appendix \ref{sec:conv_noise}.

Regardless of the exact shape in the innermost regions, it is clear that the density profiles can become shallower due to the presence of small clumps in the collapse. Moreover, the radial range that is affected correlates with the size of the clumps. Several competing mechanisms could be active here. For instance, one could think of how the small clumps induce or modify violent relaxation during collapse of the prompt cusp. As it forms, the clumps merge with each other, adding further stochastic potential fluctuations to the dynamical relaxation that is going on. When the clumps are larger, the variations in the potential are stronger. This could possibly explain the correlation with the size of the clumps. 

Another possible mechanism to explain the shallowing of the inner profile is the heating of the distribution in phase space. Random motions induced by the additional small-scale structure act as an effective velocity dispersion, which limits how much the material can be compressed in phase space. However, due to the complicated phase-space structure of the small scale medium, this does not give a single global phase space constraint, but rather a spectrum of permitted phase space densities. As we explore in Appendix \ref{sec:phase_space}, this argument can at least be used to determine the scale where the profile with noise becomes shallower than the cusp profile. 

While both mechanisms are plausible and may play a role in setting the shallowing of the profiles, it is not straightforward to quantify the relevance of each of them, and discern which is the dominant. We leave this for future work.

Lacking a reliable physical explanation to build a model upon, we provide a simple empirical fit of the maximum radius that is affected by the small-scale structure: $r_{\rm max}$. We define such a radius as the point where the density profiles deviate by $20\%$ from the power-law behaviour, as indicated by the vertical dotted lines in Figure \ref{fig:noise_dens}. Taking advantage of the correlation between this scale and the mass of the largest clumps, we fit the following empirical formula:

\begin{equation}\label{eq:empirical_fit}
    r_{\rm max} = a \left(\frac{M_{\rm max}}{M_{\rm hm}} \right)^b r_{\rm cusp},
\end{equation}

\noindent where $a$ and $b$ are free parameters, $M_{\rm max}$ is the mass of the largest clump at the time of collapse, $M_{\rm hm}$ is the half-mode mass, and $r_{\rm cusp}$ is the radius of the cusp as given by \cite{Delos:2019}\footnote{We adopt $\alpha_{\rm cal}=24$ as calibrated by \cite{Delos:2022} through all the paper, unless stated the contrary. }. The best fitting parameters are $a=1.8136$ and $b=0.4396$. In the next section we will use this empirical formula to constrain the region influenced by artificial fragments in haloes of $N$-body simulations. 

%\section{$N$-body simulations with varying cut-offs}\label{sec:Nbsims}
\section{Quantifying the effect of artificial fragments in $N$-body simulations}\label{sec:Nbsims}

Despite the issues described in detail in previous sections, $N$-body simulations can be very useful to study the formation of structures in general and first haloes in particular. In order to draw reliable conclusions, though, the regions affected by artificial fragmentation need to be discarded. In this section, we combine the ideas presented in Section \ref{sec:artfrags} with a set of $N$-body simulations (Section \ref{sec:wdm_models}), and we provide an empirical formula to identify the regions that may be affected by this numerical issue (Section \ref{sec:artfrags_Nb}). 

%In this section, we use a set of $N$-body simulations to study first halo formation, exploring a wide range of WDM cosmologies with varying initial power spectra that lead to different formation paths  and hence, potentially, to different halo structure (Section \ref{sec:wdm_models}). In order to draw reliable conclusions, we use the empirical formula obtained in Section \ref{sec:artfrags} to identify scales that may be affected by artificial fragmentation (Section \ref{sec:artfrags_Nb}). Discarding such regions, we study how the inner structure of first haloes depends on the shape and scale of the WDM cut-off (Section \ref{sec:universality}). 

\subsection{Power spectra}\label{sec:wdm_models}

\begin{figure}
    \centering
    \includegraphics[width=0.9\columnwidth]{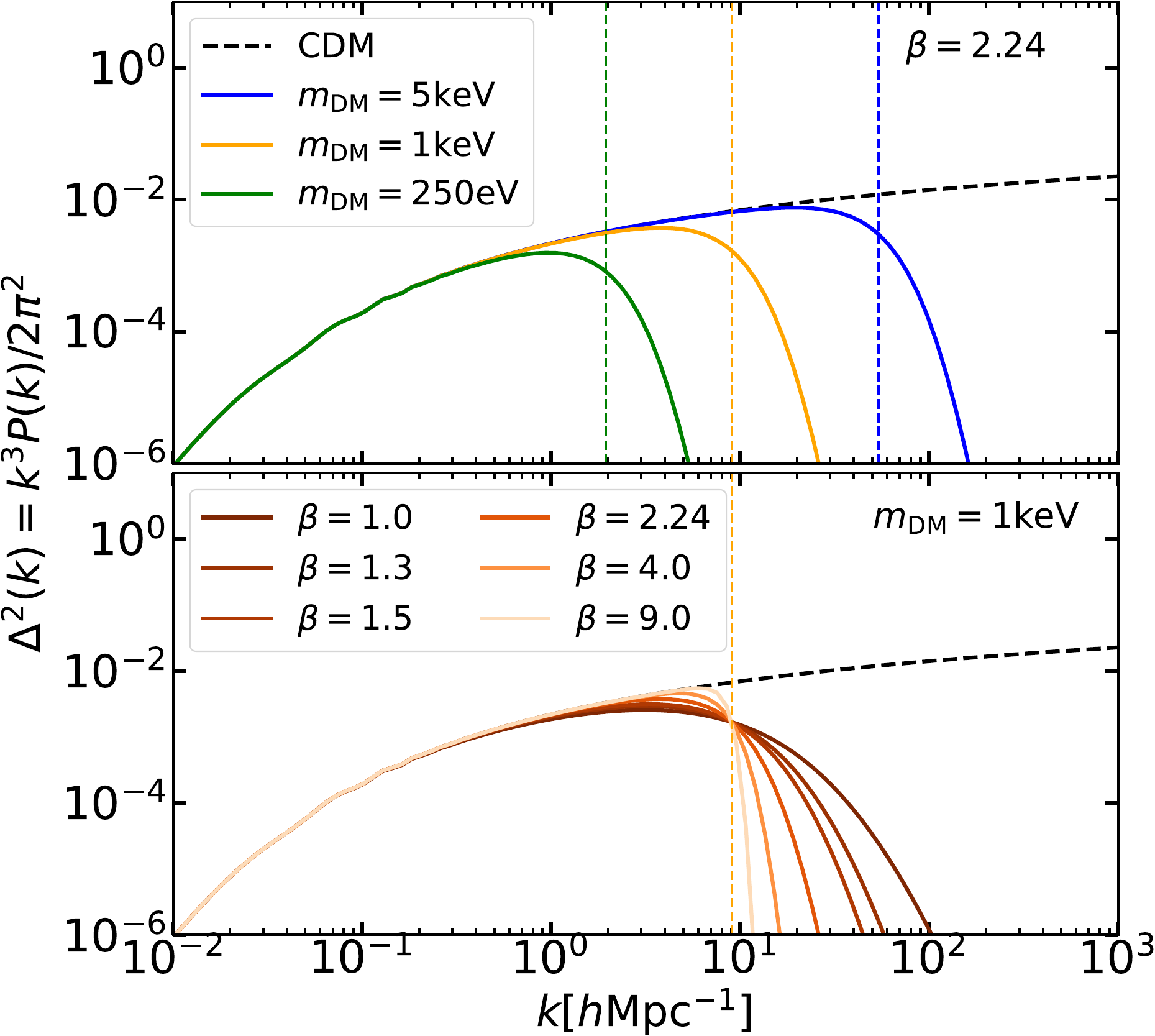}
    \caption{Dimensionless power spectra of the WDM models that we explore in this work. In the upper panel we vary the dark matter particle mass, keeping the shape of the cut-off fixed ($\beta=2.24$). In the lower panel, we vary the shape of the cut-off at fixed particle mass ($m_{\rm DM}=1\rm keV$). The vertical lines denote the half-mode scale for each spectrum. The black dashed line is the CDM power spectrum. The details of each WDM model are listed in Table \ref{table:wdm_models}.}
    \label{fig:cosmo_Nbody}
\end{figure}

The initial power spectra that we consider are displayed in Figure \ref{fig:cosmo_Nbody}. The two properties that we vary are the scale (upper row) and the shape (lower row) of the WDM transfer function, as set by $\alpha$ and $\beta$ in Equation \ref{eq:transfer}, respectively. The parameters of these cosmologies are listed in Table \ref{table:wdm_models}. The scale is set by the mass of the (thermally produced) dark matter particle, ranging from $m_{\rm DM} = 250~\rm eV$ in the hottest, to $m_{\rm DM} = 5~\rm keV$ in the coldest case we consider. This range displaces the scale of the cut-off by more than an order of magnitude. As a consequence, the masses of the smallest haloes that form in these cosmologies vary by more than 4 orders of magnitude: from $M_{\rm hm} \sim 10^{12} h^{-1} \rm M_{\odot}$ to $M_{\rm hm} \sim 6 \cdot 10^{7} h^{-1} \rm M_{\odot}$. As the slope of the power spectrum varies with $k$, changing $\alpha$ leads to different effective slopes at the cut-off. This influences the average curvature of the initial peaks, and possibly, also their collapse.

At fixed dark matter particle mass, we also  consider different shapes of the cut-off. Very steep cut-offs ($\beta \sim 9$) strongly suppress  small scale power, while in shallower cases ($\beta \sim 1$) the suppression is more progressive and extended allowing formation of smaller scale haloes below $M_\text{hm}$. In the later case, the peaks will collapse in a much more crowded environment, surrounded by other peaks of a range of masses. As is clearly visible in our simulations, by the time the haloes reach $M \sim M_{\rm hm}$, they have, on average, gone through several mergers or large mass accretion events. In the former case, the peaks collapse and get to $M \sim M_{\rm hm}$ exclusively through monolithic collapse and smooth accretion. 

Finally, for each of these set-ups, we have simulated the formation of several haloes (c.f. set 3 in Table \ref{tab:halos}), in which we keep the power spectrum fixed, but vary the formation redshift of the halo. Haloes forming at lower redshifts have lower initial peak heights, whereas those forming at high redshift have higher peak heights. 

\subsection{Modelling artificial fragmentation}\label{sec:artfrags_Nb}

\begin{figure*}
    \centering
    \includegraphics[width=\textwidth]{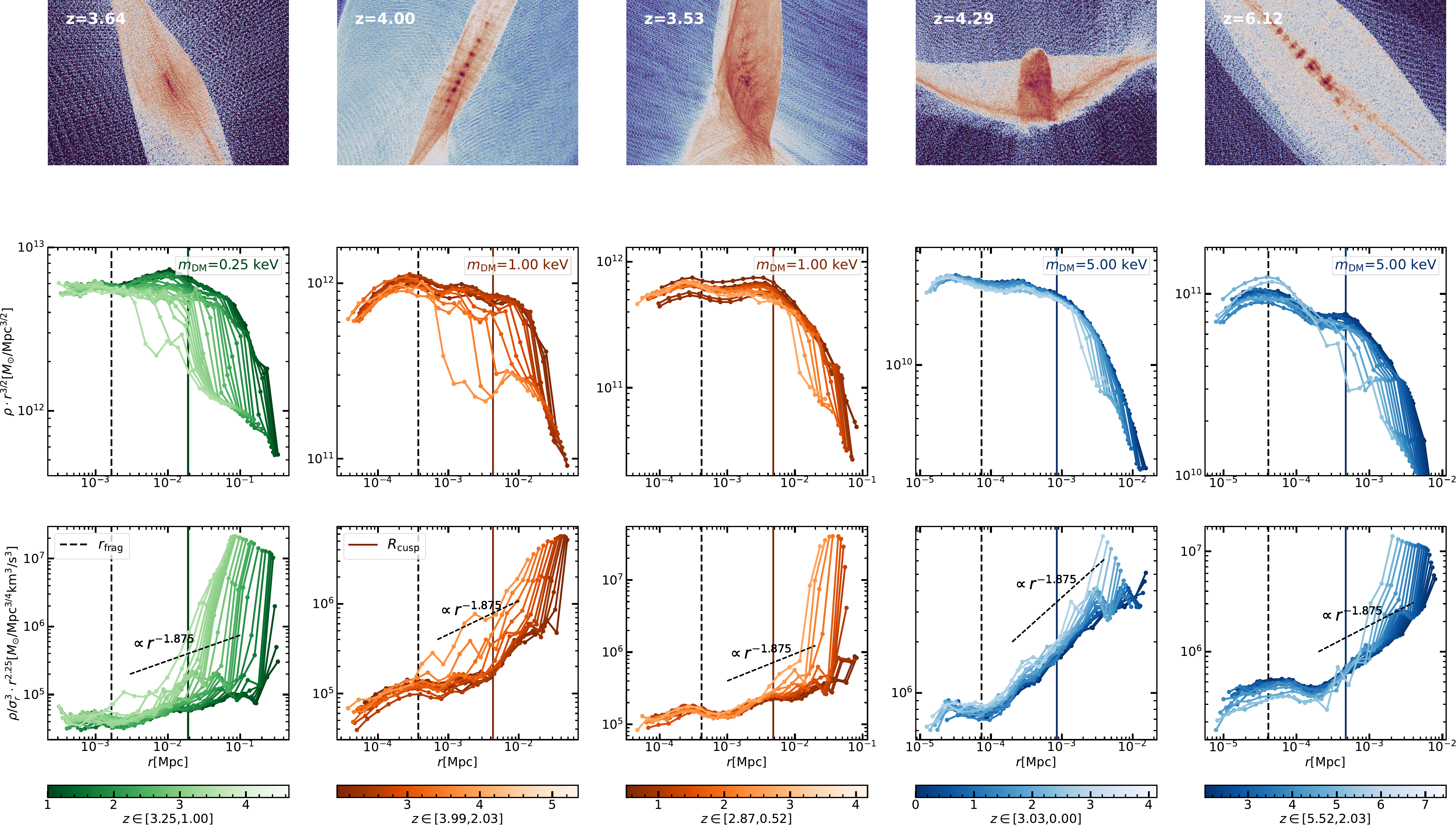}
    \caption{A selection of $N$-body simulations of the formation of first haloes. In the upper row we show the pre-collapse field, and in the middle and lower panels the density (times $r^{1.5}$) and pseudo phase-space density profiles (times $r^{2.25}$) of the resulting haloes at different redshifts. The colored vertical solid lines are the cusp radii given by the model of \protect \cite{Delos:2019}. The vertical dashed lines are a conservative estimate of the largest radius affected by the presence of artificial fragmentation in the collapse, $r_{\rm frag}$ (Equation \ref{eq:empirical_fit_Nb}).}
    \label{fig:core_frags}
\end{figure*}

As we concluded in previous sections, the inner profiles produced by $N$-body simulations cannot be trusted without question, as they may be affected by artificial fragmentation. In Figure \ref{fig:core_frags} we display a selection of haloes where we illustrate the variety of formation paths that we find in our $N$-body simulations. The upper panels depict the pre-collapse field, before the haloes form; and the middle panels represent the density profiles of the resulting haloes. The different redshifts are indicated by colors. All haloes host a cusp with $\rho \propto r^{-1.5}$. However, when the pre-collapse field is clearly fragmented (second and last column), the cusp profile becomes  shallower in the innermost region -- i.e. the cusp is destroyed due to numerical artefacts.

Notice that the effect of artificial fragmentation is localised: only the innermost region of the cusp becomes shallower. The affected region does not evolve with redshift either. Hence, the issue is to identify the radial range that is affected by artificial fragmentation and discard it when analysing $N$-body simulations, keeping only the reliable parts of the density profiles. Guided by our experiments with controlled small-scale power, we use the empirical fit of Equation \ref{eq:empirical_fit} to define a fragmentation radius $r_{\rm frag}$ within which artificial clumps may have an effect, but outside which we consider the profiles to be reliable.

The simulations considered here show several differences compared to the experiments in Section~\ref{sec:small-scale-noise}. Whereas  in the controlled simulations the pre-collapse field had an abundance of small clumps, artificial clumps appear only in filaments, and not all filaments fragment before the haloes collapse. A complete analysis should consider and model these details. Nevertheless, as a first approximation, we will consider the results of Section \ref{sec:small-scale-noise} to provide a conservative estimate of $r_{\rm frag}$. 

In our modelling we will assume:  1) that all filaments are completely fragmented before the collapse of the halo, 2) that the fragments are as big as they can get in each case, and 3) that the size of the fragments does not depend on resolution. {\it A priori} one might expect that the typical size of fragments to be several hundreds to thousands of particles. However, the dependence with resolution is very weak ($M_{\rm frag} \propto m_p^{1/3}$ according to \citealt{Wang:2007}). Furthermore, even if that scaling held at fragment formation, by the time of halo collapse, fragments may have grown further by merging (see the upper panels of Figure \ref{fig:shvsnbcollapse_5}) washing out the already weak dependence on resolution. Indeed, in our simulations we find in the worst cases that the largest fragments are $ M_{\rm frag} \sim 0.1 \%$ of the half-mode mass at the moment of halo collapse, regardless of the resolution of the simulation. To be conservative, we will assume that $M_{\rm frag} = 0.001M_{\rm hm}$. With these assumptions, $r_{\rm frag}$ reduces to a fixed fraction of the cusp radius,
\begin{equation}\label{eq:empirical_fit_Nb}
    r_{\rm frag} = a \left( 0.001 \right)^b r_{\rm cusp} = 0.087 r_{\rm cusp}.
\end{equation}
The predicted fragmentation radius are overplotted in Figure \ref{fig:core_frags} as vertical dashed lines. Of course, when the pre-collapse filament does not fragment, the fragmentation radius is overestimated (first, third, and fourth columns). However, when the pre-collapse filament does fragment, the fragmentation radius lies where the profiles start to deviate from the steep cusp (second and fifth column). This is remarkable, considering that the empirical formula was not fitted to the $N$-body simulations with artificial fragmentation, but to the controlled experiments with small scale power of Section \ref{sec:small-scale-noise}. The fact that the fit performs well to predict the extent of the modified region due to artificial fragments indicates that the details of the small-scale structure are not relevant on setting this scale.

\subsection{Numerical effects on halo profiles}

\begin{figure}
    \centering
    \includegraphics[width=0.9 \columnwidth]
    {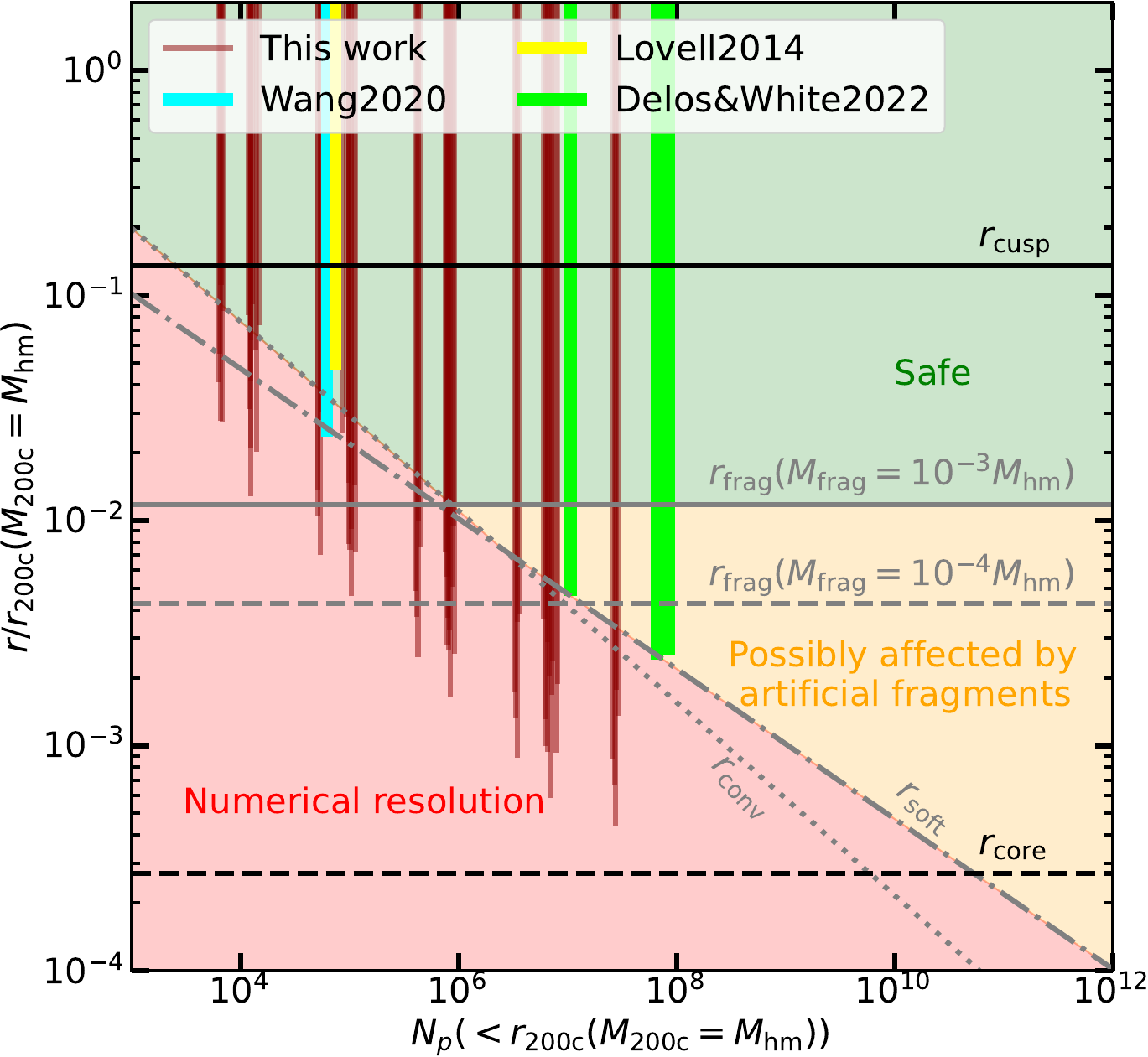}
    \caption{Radial range of the density profiles of the first haloes at the half-mode mass in $N$-body simulations, as a function of resolution. The red area shows the region that can not be trusted due to numerical resolution (2-body relaxation processes and force softening). The orange area is the dangerous area, that could possibly be affected by artificial fragments. In the green region, the profiles are reliable. We overplot the cusp radius predicted from the characteristics of the initial peak \protect\citep{Delos:2019}, and core radius ($\sim r_{\rm cusp} / 500$) expected due to the initial velocity dispersion of dark matter. The vertical lines represent our simulations, until the radius that encloses 100 particles. Furthermore, we make an orientative comparison with some profiles in the literature: the highest resolution level of \protect\cite{Wang:2020}, the smallest subhalo of \protect\cite{Lovell:2014}, and all haloes from \protect\cite{Delos:2022}.}
    \label{fig:conv}
\end{figure}

Artificial fragmentation causes an additional numerical shallowing of the profiles of first haloes, on top of the well-known effects of 2-body relaxation and force softening. Yet, these effects are relevant in different regimes, and it can be useful to know how they relate to each other to understand the dominant effect in each configuration. In Figure \ref{fig:conv}, we show the radial range of halo profiles in units of $r_{\rm 200c}$ ($r/r_{\rm 200c}$) as a function of the number of particles that have been used to resolve each halo ($N_p(<r_{\rm 200c}$)), for haloes at the half-mode mass, $M_{\rm 200c}=M_{\rm hm}$. The grey lines denote the radii affected by different numerical effects: two-body relaxation ($r_{\rm conv}$, given by \citealt{Power:2003}), force softening ($r_{\rm soft} = 3 \epsilon$, where $\epsilon$ is the Plummer-equivalent force softening parameter), and artificial fragmentation ($r_{\rm frag}$, Equation \ref{eq:empirical_fit_Nb}). We also overplot as black lines the expected outer radius of the prompt cusp ($r_{\rm cusp}$, given by \citealt{Delos:2019}) and the core radius coming from the initial velocity dispersion of dark matter ($r_{\rm core} \sim r_{\rm cusp}/500$). All the relations were derived with our set of $N$-body simulations at different resolutions. We display them as brown vertical lines, plotting down to the radius that encloses 100 particles. The numerical parameters of the resolution levels are listed in Table \ref{tab:params_zooms}. Of course, for each of the aforementioned radii, we have obtained a distribution rather than a single line. However, these distributions are fairly tight and do not change the message of the plot. For clarity, we decided to show only the mean value and ignore the scatter around it.

In the green region, over the grey lines, the outputs of $N$-body simulations can be trusted. Depending on the resolution, different numerical effects are dominant and define the extension of this region. At low resolution, $N_p(<r_{\rm 200c}) < 10^6$, the limiting factor is the numerical resolution of our simulations (red region) -- i.e. the softening and 2-body relaxation. At higher resolution, $N_p(<r_{\rm 200c}) > 10^6$, the artificial fragmentation becomes non-negligible, creating the orange `dangerous' region. It is worth emphasizing that this is a conservative prediction: the orange region is not necessarily inaccessible to $N$-body simulations. If the pre-collapse density field does not fragment, or if the fragments are much smaller than what we assumed, the results of the $N$-body simulations can be trusted at smaller radii. This is illustrated by the dashed grey line: if the fragments were an order of magnitude smaller than what we assumed, the profiles would be unaffected at almost three times smaller radii. 

Yet, notice that while the convergence and softening radii decrease fast with resolution, the fragmentation radius does not. Thus, as we increase the resolution of our simulations, the possible presence of artificial fragmentation becomes increasingly problematic. In particular, providing a reliable $N$-body simulation that can resolve the scale that is relevant for the formation of a phase space core, seems difficult. However, a sheet based simulation with a refinement plus phase space reconstruction step close to collapse (similar to Section \ref{sec:halos}) could be viable. 

One could also think in the opposite direction: what would happen if the fragmentation is more severe than we assumed? Could the fragments completely erase the cusp? Following Equation \ref{eq:empirical_fit}, we know that for $r_{\rm frag} \sim 0.5-1 r_{\rm cusp}$, the fragments would need to be $5-20 \%$ of the half-mode mass, which is extremely unlikely.

Finally, it is interesting to observe the region where previous simulations lie in this plane. We overplot the haloes at the half-mode mass simulated in \cite{Lovell:2014,Wang:2020,Delos:2022} as yellow, blue, and green lines, respectively. We display the lines until the smallest radius that each paper considers trustworthy.\footnote{Notice that we do not need to know the $r_{\rm cusp}$ value for their haloes: it is enough to know $r_{\rm 200c}$, $r_{\rm min}$, $M_{\rm 200c}$ and dark matter particle mass.} As expected, the lines stop where the region affected by numerical resolution starts.

Some published papers claim that haloes at the half-mode mass are in agreement with a NFW profile \citep{Lovell:2014,Wang:2020}. Figure 2 of \cite{Wang:2020} shows a profile for haloes at the half-mode mass that never becomes shallower than -1.5 in the resolved range and does not allow to distinguish clearly between NFW and prompt cups profiles for such haloes. We speculate that \cite{Wang:2020} would also have measured a steeper -1.5 profile if they had reached higher resolutions in their most zoomed simulations. 

Recent simulations presented in \cite{Delos:2022} resolve more than an order of magnitude within $r_{\rm cusp}$. In agreement with our results, they find that those haloes form a steep cusp. Nonetheless, they do not see any evident shallowing on the profiles even if at least some of their filaments do fragment (see Figure 5 of \cite{Delos:2022}). This is not necessarily in tension with our findings. Recall that the fragmentation radius depends on the mass of the fragments: the ``dangerous region'' of Figure \ref{fig:conv} is derived with the conservative assumption of $M_{\rm frag}=10^{-3}M_{\rm hm}$ (solid grey line). If the fragments in \cite{Delos:2022} were smaller by an order of magnitude (dashed grey line), we would not have expected to see any shallowing in the profiles of \cite{Delos:2022}, at their current resolution. Moreover, in the highest resolution haloes of \cite{Delos:2022}, they see a tendency of the profiles to shallow in the innermost region. We speculate that this could be due to artificial fragments. 

All in all, the findings of previous studies are overall consistent with our results. In order to make a more quantitative comparison, one would have to look at the simulations and the pre-collapse density fields one by one.

%\subsection{Properties of the steep cusp}\label{sec:universality}

\section{Properties of the first generation of haloes}\label{sec:universality}

In this section, we study first halo formation exploring a wide range of WDM cosmologies with varying initial power spectra that lead to different formation paths and hence, potentially, to different halo structure (Section \ref{sec:wdm_models}). Discarding the regions that are potentially affected by artificial fragments, we test how the inner structure of first haloes depends on the shape and scale of the WDM cut-off, as well as formation redshift. In particular, we pay attention to three properties: slope of the density profile (Section \ref{sec:slope-cusp}), amplitude of the density profile (Section \ref{sec:ampl-cusp}), and pseudo-phase-space density profiles (Section \ref{sec:ph-dens-pr}). 

\subsection{Slope of the cusp}\label{sec:slope-cusp}

\begin{figure}
    \centering
    \includegraphics[width=0.9 \columnwidth]{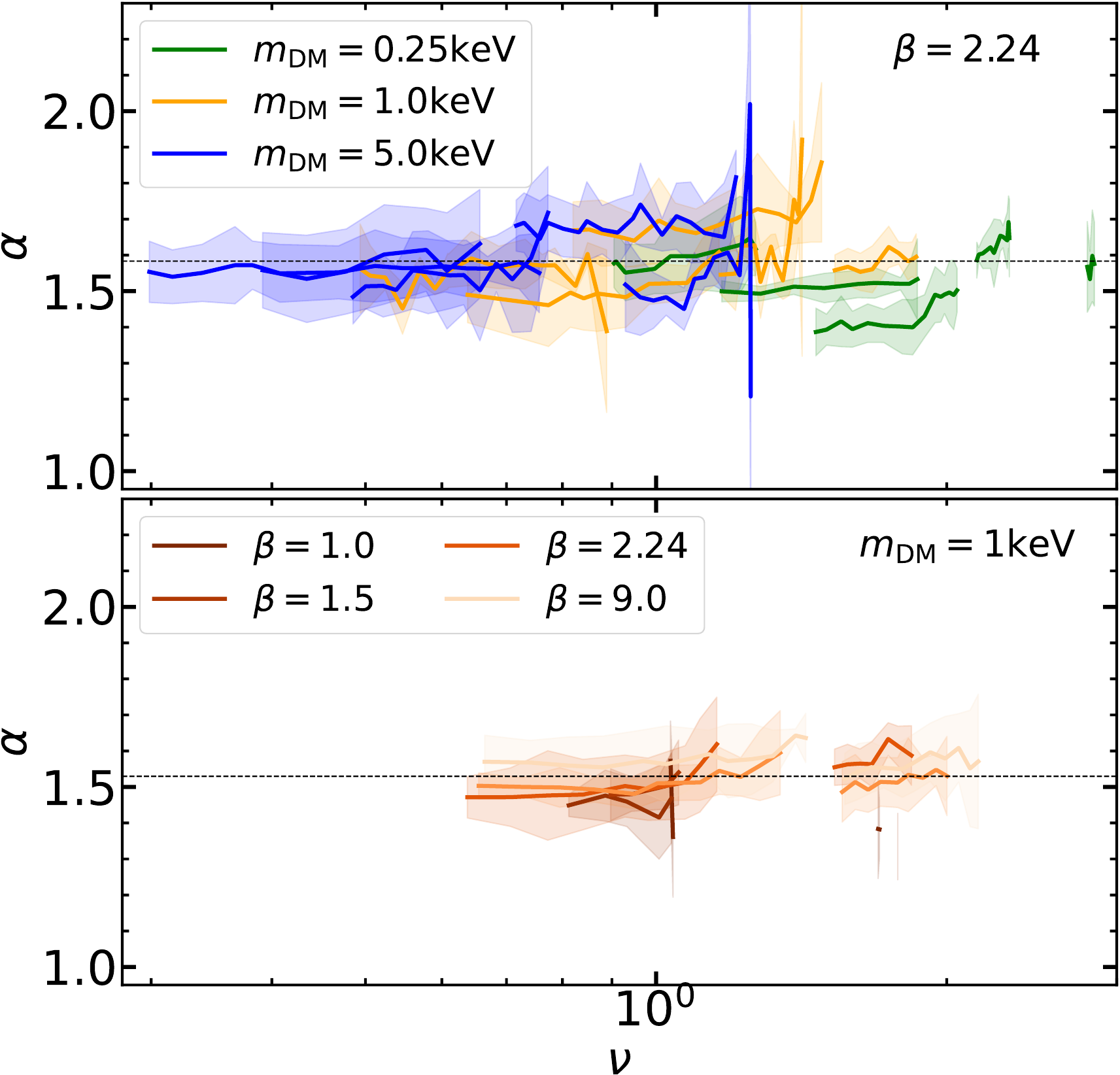}
    \caption{Distribution of the inner slope $\alpha$ of the haloes in the $N$-body set 3 as a function of peak height $\nu$. Each line corresponds to the average slope of the cusp of a halo, from the moment it forms until it reaches the half-mode mass. The shaded region denotes the uncertainty on the measurement of $\alpha$ (see text for details). The range of radii that we use to compute the average is $r_{\rm min} < r < r_{\rm cusp}/2$, where $r_{\rm min} = \max [r_{\rm soft},r_{\rm conv},r_{\rm frag}]$.
    In the upper row, we vary the warmth of the dark matter as well as the redshift of formation of the halo, keeping fixed the shape of the cut-off. In the lower row, we keep fixed the warmth, and we change the shape of the cut-off. The dashed horizontal line is the mean of the slopes in each panel.}
    \label{fig:alpha_nu}
\end{figure}

In Figure \ref{fig:alpha_nu} we show the distribution of the slope of the cusps for all the haloes of our $N$-body set. %The outcome of the third set of simulations is summarized in Figure \ref{fig:alpha_nu}. 
The slope of the inner region of the haloes, $\alpha$, is displayed as a function of peak height, $\nu = \delta_c / \sigma(M)$, where $\delta_c$ is the critical density for collapse and $\sigma(M)$ is the mass variance, $M$ being halo mass. Each line represents the average slope of a halo at different redshifts. The shaded region is the standard deviation of the different values of the slope measured at different radii, it denotes somewhat the uncertainty in the measurement of $\alpha$. The range we use to measure the slope is defined by $r_{\rm min} < r < r_{\rm cusp}/2$. The cusp radius is set by the model of \cite{Delos:2019}, and $r_{\rm min} = \max [ r_{\rm conv}, r_{\rm soft}, r_{\rm frag} ]$.

The upper panel of Figure~\ref{fig:alpha_nu} shows the distribution of the slopes varying the scale of the cut-off. There is not any evident dependence as we move from low to high $\nu$ values: all the haloes lie somewhere in the range $\alpha \in {[-1.4, -1.65 ]}$, peaking at $\sim -1.58$, slightly steeper than the usually quoted $-1.5$. This statement holds when we vary the peak height by changing the scale of the cut-off (keeping the formation redshift fixed), and when we change the formation redshift (at fixed cut-off scale) Simultaneously, we changed the slope of the power spectrum at the cut-off, and in consequence, the curvature of the initial peaks too. Besides, the fact that the lines are horizontal rather than inclined, tells us about the stability of the inner structure of these haloes: once formed, it does not evolve.  In the lower panel we show the re-simulations of two haloes varying the shape of the cut-off, at fixed scale: the lines fall on top of each other, in agreement with previous results. 

We confirm that the formation of a steep cusp $\propto -1.5$ is a general result of the gravitational collapse from the smooth field, rather than being specific to any particular configuration. This picture holds in all the explored scenarios, in which we have varied formation redshifts, warmth of dark matter or shapes of the cut-off. Within the four orders of magnitude in mass that we have tested, no evolution was found. Furthermore, if one adds the results of previous works that have found steep cusps in the Earth-mass haloes of $m_{\rm DM} = 100 \rm GeV$ neutralino dark matter scenarios \citep{Diemand:2005,Ishiyama:2014,Angulo:2017}, the mass range spans 18 orders of magnitude. This means that, in parallel to the universality of the NFW profile found in CDM cosmologies, the prompt cusp is also universal for haloes forming at the cut-off of the power spectrum.

\subsection{Amplitude of the cusp}\label{sec:ampl-cusp}

\begin{figure}
    \centering
    \includegraphics[width=0.9 \columnwidth]{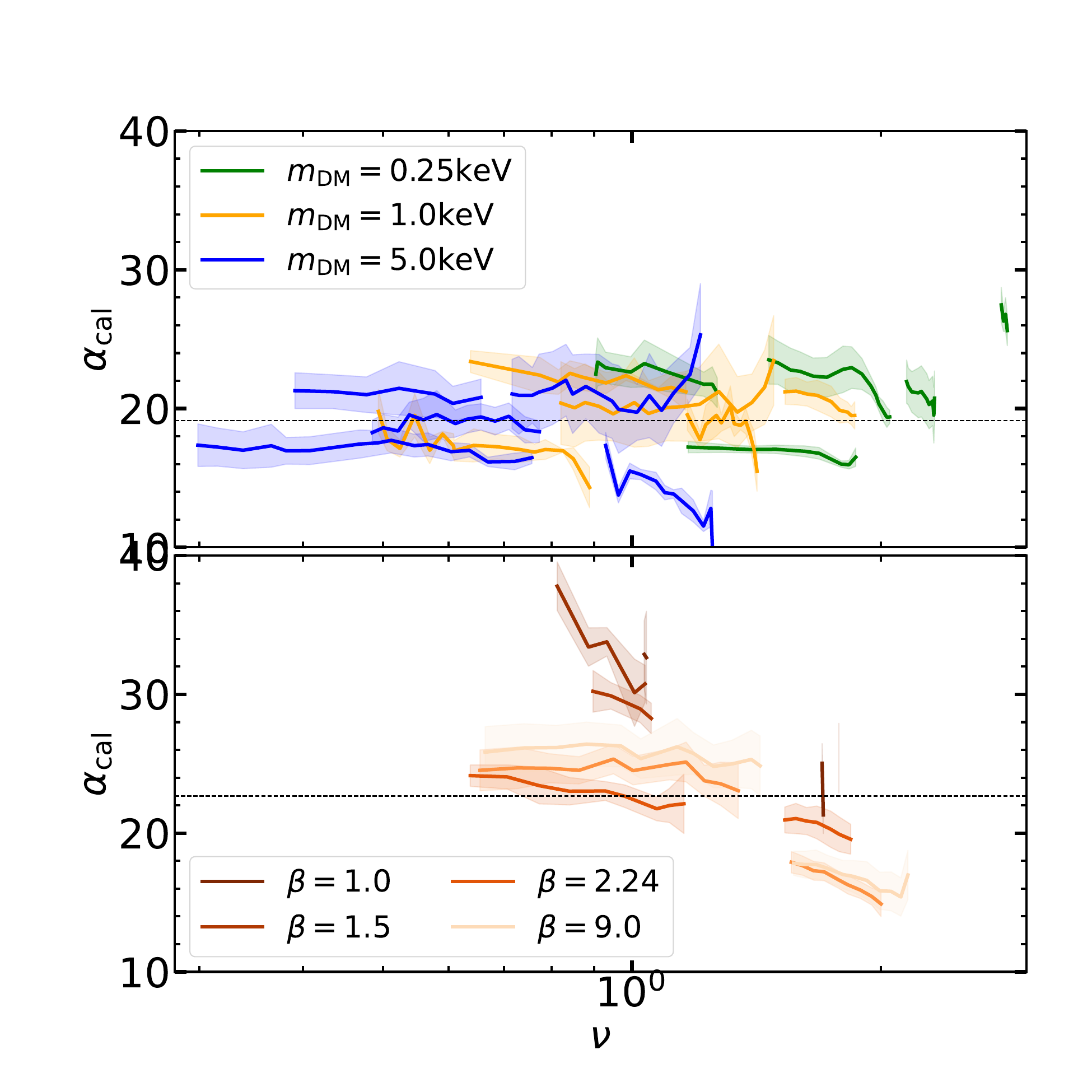}
    \caption{Distribution of the normalization constant $\alpha_{\rm cal}$ of the model of \protect\citep{Delos:2019}, calibrated on the haloes of our $N$-body set, as a function of the peak height $\nu$. Each line corresponds to the value derived from a halo at different redshifts, from the moment it forms until it reaches the half-mode mass. The shaded region represents the uncertainty in $\alpha_{\rm call}$ (see text for details). The range of radii that we use is $r_{\rm min} < r < 0.13r_{\rm 200c}$, where $r_{\rm min} = \max [r_{\rm soft},r_{\rm conv},r_{\rm frag}]$.
    In the upper row, we vary the warmth of the dark matter as well as the redshift of formation of the halo, keeping fixed the shape of the cut-off. In the lower row, we keep fixed the warmth, and we change the shape of the cut-off.}
    \label{fig:alphacall_nu}
\end{figure}

With the existence of these prompt cusps in the first haloes settled, models that predict their characteristics are very useful. For instance, the model presented in \cite{Delos:2019} has been used to put new constraints on dark matter, re-analyzing the $\gamma$-ray signal excess coming from the Galactic center in presence of these cusps \citep{Delos:2022a, Stucker:2023}. However, this model was never tested extensively on a wide range of possible cut-offs. In particular, it has a free-parameter, $\alpha_{\rm cal}$, that determines the amplitude of the cusps,
\begin{equation}
 \alpha_{\rm cal} = A / \rho_b a_{\rm ec}^{3/2} R^{-3/2},
\end{equation}

\noindent where $a_{\rm ec}$ is the time of collapse given by the ellipsoidal collapse model, $\rho_b$ is the mean density of the universe at $a_{\rm ec}$, $R \equiv |\delta / \nabla^2\delta | ^{1/2}$ is associated with the curvature of the initial density field $\delta$, and $A$ is the measured amplitude of the density profile $\rho = A r^{-3/2}$. Previous work has found $\alpha_{\rm cal} \sim 24$, to $\sim 20\%$ accuracy \citep{Delos:2022}. For us, it is straightforward to test whether this assumption holds for all the transfer functions that we have explored. 

In Figure \ref{fig:alphacall_nu} we present the results of such a test. As in Figure \ref{fig:alpha_nu}, we show the $\alpha_{\rm cal}$ value obtained for each halo at different redshifts, as a function of $\nu$. The shaded regions denote the uncertainty on the measurement of $\alpha_{\rm cal}$, summing in quadrature the uncertainty on the measurement of the amplitude $A$ at different radii, and the error on the determination of $R$. The fit to $\alpha_{\rm cal}$ is done for radii $r_{\rm min} < r < r_{\rm max}$, where $r_{\rm min} = \max [ r_{\rm conv}, r_{\rm soft}, r_{\rm frag} ]$, and $r_{\rm max}=0.13 r_{\rm 200c}$. The latter has been empirically found to capture relatively well the extent of the cusp.

Overall, the results seem to be roughly consistent with $\alpha_{\rm cal}  = 24$ at $\sim 20\%$ of accuracy, even if our distribution peaks at $\alpha_{\rm cal} = 19.1$. As the annihilation radiation goes like $J \propto A^2 \propto \alpha_{\rm cal}^2$ \citep{Stucker:2023}, this changes the cusp contribution to the annihilation radiation by a factor of 1.6. Regarding the different cut-off shapes (lower panel), we see a larger scatter, and even an evolution of $\alpha_{\rm cal}$ mainly in low $\beta$ values. This is related to the amount of small scale structure in low $\beta$ cosmologies. While the physical meaning of $\alpha_{\rm cal}$ is related to the mass that collapses monolithically into the cusp from the initial peak, in these cases different peaks merge, or large amounts of mass accretion happen soon after the monolithic collapse. This mass injection boosts the amplitude of the cusp $A$, and results in an increase of the calibrated $\alpha_{\rm cal}$\footnote{Even if not shown here, when strong mergers happen, we find that the cusp can be even destroyed in the very early stages of the halo. This is in agreement with previous work \citep{Angulo:2017, Ogiya:2016, Ogiya:2018}.}. Thus, one has to be careful when using this model in transfer functions with low $\beta$, where the cut-off is very extended, and the small scales are not suppressed so fast. 

\subsection{Pseudo-phase-space density profile of the cusp}\label{sec:ph-dens-pr}

Another interesting property to look at are pseudo-phase-space density profiles. $N$-body simulations show that for standard CDM haloes, this quantity is a power-law to a good approximation $ \rho / \sigma^3 \propto r^{\chi}, \chi \sim -1.875$, even though both the density profiles and the velocity dispersion profiles are broken power laws \citep{Taylor:2001,Ludlow:2010,Ludlow:2011,Colombi:2021}. This can be explained if one takes into account that density profiles have evolving power-law index \citep{Dehnen:2005,Ludlow:2011}. It is not clear {\it a priori} whether this scaling should extend down into prompt cusps, since they clearly form by a qualitatively different process than later generations of haloes. 

For a power-law density profile, the expected behaviour of the pseudo phase-space density profile can be easily estimated. Under the assumption of spherical symmetry and isotropic velocity dispersion, the Jeans equation is
\begin{equation}
    \frac{d(\rho \sigma_r^2)}{dr} = -\rho(r) \frac{GM(r)}{r^2},
\end{equation}
where $\sigma_r$ is the radial velocity dispersion. Introducing a power-law density profile $\rho = A r^{-\alpha}$, this gives
\begin{equation}
    \sigma_r =  \left[\frac{2\pi A G}{(3-\alpha)(\alpha-1)}\right]^{1/2}r^{(2-\alpha)/2}.
\end{equation}
Combining the two previous results, we get
\begin{equation}
    Q_r = \frac{\rho}{\sigma_r^3} = \left[\frac{(3-\alpha)(\alpha-1)}{2 \pi G}\right]^{3/2} A^{-1/2} r^{\alpha/2-3}.
\end{equation}

Prompt cusps have isotropic velocity dispersions and approximately power-law density profiles $\rho \propto r^{-1.5}$ \citep[see][and the discussion above]{Delos:2022}). Thus, if both properties were to extend over a large dynamic range, Jeans equation requires a power-law pseudo-phase-space density profile with $\chi = -2.25$.

In the lower panels of Figure \ref{fig:core_frags} we display the pseudo phase-space density profiles multiplied by $r^{2.25}$. For a wide range in radii around $R_{\rm cusp}$, the profiles are proportional to a power-law with index $\chi \sim -1.875$, even if there are systematic deviations (note the greatly expanded vertical axes in these plots). Interestingly, in the innermost region, the pseudo phase-space density profiles become steeper, and indeed although the slope within the nominal prompt cusp radius appears to vary, in several cases its mean value is close to $\chi = -2.25$ (in agreement with the findings of \cite{Ishiyama:2010}). However, we would require higher resolution simulations to judge this with certainty. A likely reason that the $Q \propto r^{-2.25}$ holds for a much smaller range than the $\rho \propto r^{-1.5}$, is that the velocity dispersion at a given radius not only depends on the force field at that radius, but also on those from $\sim 2-3$ larger radii. Furthermore, notice that the material that collapsed monolithically to the halo most likely resides at $r<<R_{\rm cusp}$, while the rest of the halo was built up by smooth accretion. 

As shown by \cite{Delos:2019} and \cite{Delos:2022} and verified for our own simulations in Figure~\ref{fig:alphacall_nu} the normalisation of cusp density profiles, and hence also of their pseudo-phase-space density profiles is set purely by the properties of the immediate neighborhood of the peak in the linear density field from which they form. It is thus intriguing that we see no indication of an offset or feature in Figure~\ref{fig:core_frags} at the boundary between the the prompt cusp and the outer main body of our haloes. Finally, as already noted by \cite{Ludlow:2010} in the outskirts, at $r\gg R_{\rm cusp}$ the fall-off of the the pseudo-phase-space becomes shallower.

\section{Summary and conclusions}\label{sec:concl}

In this paper, we have carried out an ensemble of numerical simulations to study the first generation of haloes that form at the free-streaming scale of dark matter. 

For this, we have extended the sheet-based numerical techniques presented in \cite{Stucker:2020}, which allows us, for the first time, to carry out high-resolution simulations devoid of the artificial fragmentation that plague standard $N$-body simulations (Figure \ref{fig:sheet_example}). Specifically, we have developed a refinement technique that provides a zoom-in capability in the quasi-linear regime, reducing dramatically the computational resources required (Table \ref{tab:sh_nb_num_CPUh}).

Our main findings can be summarised as follows:

\begin{itemize}
    \item Our sheet simulations do find that primordial density peaks indeed collapse into steep prompt cusps with power-law density profiles, $\rho\propto r^{-1.5}$ (Figure \ref{fig:shvsnbdensity}). This confirms previous findings in the literature and eliminate doubts that they could be the result of numerical inaccuracies.
    \item By comparing our sheet simulations with $N$-body counterparts, we suggest that if artificial fragmentation occurs after a halo collapses, then steep prompt cusps also form in $N$-body simulations. In contrast, if fragmentation occurs before the collapse, then the inner density may be significantly reduced (Figure \ref{fig:shvsnbdensity}).
    \item Using a suite of simulations with controlled small-scale noise, we were able to identify the scales that could be affected by artificial fragmentation (Figure \ref{fig:conv}). This scale depends weakly on resolution, thus, for haloes with more than $10^8$ particles, the effect of artificial fragmentation might be more important than that of force softening and 2-body relaxation. 
    \item Once we identified the range of scales where $N$-body simulations are robust, we carried out a suite of $N$-body simulations of the first haloes in a variety of WDM cosmologies. Our simulations showed that, regardless of formation time or details of the cut-off in the initial power spectrum, first haloes always host a $\rho\propto r^{-1.5}$ prompt cusp in their innermost regions. 
    \item Additionally, we find that the pseudo phase-space-density profiles of prompt cusps are consistent with $Q \propto r^{-2.25}$ in the innermost regions, transitioning to $Q \propto r^{-1.875}$ at intermediate radii. Although this behaviour is not completely understood yet, in the halo center it agrees with the expected scaling for prompt cusps. 
\end{itemize}

In light of these results, we consider that the formation of prompt cusps at the moment of collapse of density peaks is now firmly established, beyond any lingering doubts about the validity of the numerical simulations which have found them.  Remarkably, this statement holds over all the cosmologies we have considered, from extremely light ($m_{\rm DM} = 250 \rm eV$) to heavier ($m_{\rm DM} = 5 \rm keV$) dark matter particles, where the mass of first haloes range from $M \sim 6 \cdot 10^7$ to $10^{12} h^{-1}\rm M_{\odot}$. 

Including results from previous work on neutralino dark matter \citep{Ishiyama:2014,Angulo:2017}, where the first haloes are Earth-mass objects, prompt cusps have been found to form in the first generation of haloes over 18 orders of magnitude in halo mass. Therefore, the formation of steep prompt cusps appears to be a universal consequence of gravitational collapse, and it is insensitive to specific details of the initial density peak and its surroundings \citep[see also][]{Delos:2022a}.

Prompt cusps appear likely to survive as larger haloes build up through smooth accretion and hierarchical merging. Although some uncertainty remains on exactly how long they will survive once accreted into galaxy and cluster-scale haloes today, their high densities mean that they are resistant to disruption everywhere except the inner regions of galaxies \citep[e.g.][]{Stucker:2023}. This might create additional observational venues to constrain warm dark matter \citep{Delos:2023}. Additionally, surviving prompt cusps likely dominate possible self-annihilation signals from dark matter in haloes \citep{Ishiyama:2010,Delos:2022a,Stucker:2023}.

To firmly establish and quantify the observational consequences of prompt cusps, it is now important to understand their very inner regions, which are crucial both for the likelihood of tidal disruption and for the expected self-annihilation signal. At the very centre of cusps, we expect phase-space constraints to limit the density and thus create a core \citep{Tremaine:1979,Stucker:2023}. Since the phase-space core is expected to appear at $r_{\rm core} \sim 0.002r_{\rm cusp}$ \citep{Stucker:2023}, which corresponds to $\sim 0.022 r_{\rm frag}$ (see Figure \ref{fig:conv}), artificial fragmentation effects will make it difficult to reach with standard $N$-body techniques. Additionally, shot-noise from sampling the intrinsic dark matter velocity distribution gives rise to unacceptable levels of small-scale noise in $N$-body simulations  \citep{Power:2016,Leo:2017}. In contrast, our sheet simulations and refined zoom-in technique could offer a robust numerical method if the refinement is extended to velocity space. We plan to investigate this in the future. Arguably, this would be the only path to numerically study the properties and implications of the very inner regions of prompt cusps, helping to robustly place constrains on the nature of dark matter.

\section*{Acknowledgements}
The project that gave rise to these results received the support of a fellowship from ”la Caixa” Foundation (ID 100010434). The fellowship code is LCF/BQ/DR21/11880028. LO, REA and JS acknowledge the support of the ERC-StG number 716151
(BACCO). REA acknowledges the support of the Project of excellence Prometeo/2020/085 from
the Conselleria d’Innovació, Universitats, Ciència i Societat Digital de la Generalitat Valenciana, and of the project PID2021- 128338NB-I00 from the Spanish Ministry of Science. The authors acknowledge the computer resources at MareNostrum and the technical support provided by Barcelona Supercomputing Center (RES-AECT-2022-1-0027).

%%%%%%%%%%%%%%%%%%%%%%%%%%%%%%%%%%%%%%%%%%%%%%%%%%
\section*{Data Availability}
 
The data underlying this article will be shared on reasonable request to the corresponding author.



\bibliographystyle{mnras}
\bibliography{main} % if your bibtex file is called example.bib

%\begin{thebibliography}{99}
%\end{thebibliography}

%%%%%%%%%%%%%%%%%%%%%%%%%%%%%%%%%%%%%%%%%%%%%%%%%%

%%%%%%%%%%%%%%%%% APPENDICES %%%%%%%%%%%%%%%%%%%%%

\appendix

% \section{Resolution dependence of artificial fragmentation}\label{sec:res_dep_fragmentation_nb}

% \begin{figure}
%     \centering
%     \includegraphics[width=\columnwidth]{figures/warmsim1_obj4_res_dep_frag_z4.00_nb.pdf}
%     \caption{Resolution dependent fragmentation of a pre-collapsed filament in an $N$-body simulation.}
%     \label{fig:res_dep_fragmentation_nb}
% \end{figure}

\section{Convergence}\label{sec:convergence}
\begin{table*}
\centering
    \begin{tabular}{|c c c c c c c | }
\hline
     & $m_{\rm DM} [\rm keV]$& level & parent level & $m_p [h^{-1}\rm M_{\odot}]$ & $\epsilon [h^{-1} \rm kpc]$ & $n_p(M_{\rm hm})$\\
\hline
    025-parent & 0.25& 7&-& $1.2 \cdot 10^9$& 5.0& $1.12 \cdot 10^3$\\
    025-L1 & 0.25& 8& 7& $1.49 \cdot 10^8$ & 2.5& $9.0 \cdot 10^3$\\
    025-L2 & 0.25& 9& 7& $1.87 \cdot 10^7$ & 1.25& $7.2 \cdot 10^4$ \\
    025-L3 & 0.25& 10& 7& $2.34 \cdot 10^6$ & $6.25 \cdot 10^{-1}$& $5.78 \cdot 10^5$\\
    025-L4 & 0.25& 11& 7& $2.92 \cdot 10^5$& $3.12 \cdot 10^{-1}$& $4.62 \cdot 10^6$\\
    025-L5 & 0.25& 12& 7& $3.65 \cdot 10^4$& $1.56 \cdot 10^{-1}$& $3.69 \cdot 10^7$\\
    \hline
    1-parent & 1.0 &9&-& $1.87 \cdot 10^7$& 1.25& $7.11 \cdot 10^2$\\
    1-L1 & 1.0 &10& 9& $2.34 \cdot 10^6$& $6.25 \cdot 10^{-1}$ & $5.68 \cdot 10^3$\\
    1-L2 & 1.0 &11& 9& $2.92 \cdot 10^5$ & $3.12 \cdot 10^{-1}$& $4.55 \cdot 10^4$\\
    1-L3 & 1.0 &12& 9& $3.65 \cdot 10^4$& $1.56 \cdot 10^{-1}$& $3.64 \cdot 10^5$\\
    1-L4 & 1.0 &13& 9& $4.56 \cdot 10^3$& $7.81 \cdot 10^{-2}$& $2.91 \cdot 10^6$\\
    1-L5 & 1.0 &14& 9& $5.71 \cdot 10^2$& $3.91 \cdot 10^{-2}$& $2.33 \cdot 10^7$\\
    1-L6 & 1.0 &15& 9& $71.33$ & $1.95 \cdot 10^{-2}$& $1.86 \cdot 10^8$ \\
    \hline
    5-parent & 5.0 &12& 9& $3.65 \cdot 10^4$ & $1.56 \cdot 10^{-1}$& $1.7 \cdot 10^3$\\
    5-L1 & 5.0 &13& 12&$4.56 \cdot 10^3$ & $7.81 \cdot 10^{-2}$ & $1.38 \cdot 10^4$\\
    5-L2 & 5.0 &14& 12& $5.71 \cdot 10^2$& $3.91 \cdot 10^{-2}$& $1.1 \cdot 10^5$\\
    5-L3 & 5.0 &15& 12& $71.33$ & $1.95 \cdot 10^{-2}$& $8.8 \cdot 10^5$\\
    5-L4 & 5.0 &16& 12& $8.91$ & $9.76 \cdot 10^{-3}$ & $7.04 \cdot 10^6$\\

    \end{tabular}
    \caption{Numerical parameters of $N$-body parent simulations and zoom-in levels. For each dark matter mass ($m_{\rm DM}$), we list the resolution of the zoom and the parent simulation, as well as the particle mass ($m_p$) and softening length ($\epsilon$) of the highest resolution particles. In the first column we state the name of each resolution level, linking the dark matter mass and the zoom-in level. As an indication of the resolution, we also state the number of particles that resolves the half-mode mass in each case ($n_p(M_{\rm hm})$).}
   \label{tab:params_zooms}
\end{table*}

In this appendix we test the convergence of halo density profiles with increasing zoom-in level. We divide the appendix in three parts, one for each simulation type presented in the main body of the paper: $N$-body simulations of WDM (Section \ref{sec:halos} and \ref{sec:Nbsims}), refined S+R simulations of WDM (Section \ref{sec:halos}), and $N$-body simulations with controlled small-scale power (Section \ref{sec:artfrags}). 

\subsection{$N$-body zoom-in}\label{sec:conv_Nb}
\begin{figure*}
    \centering
    \includegraphics[width=1.0\textwidth]{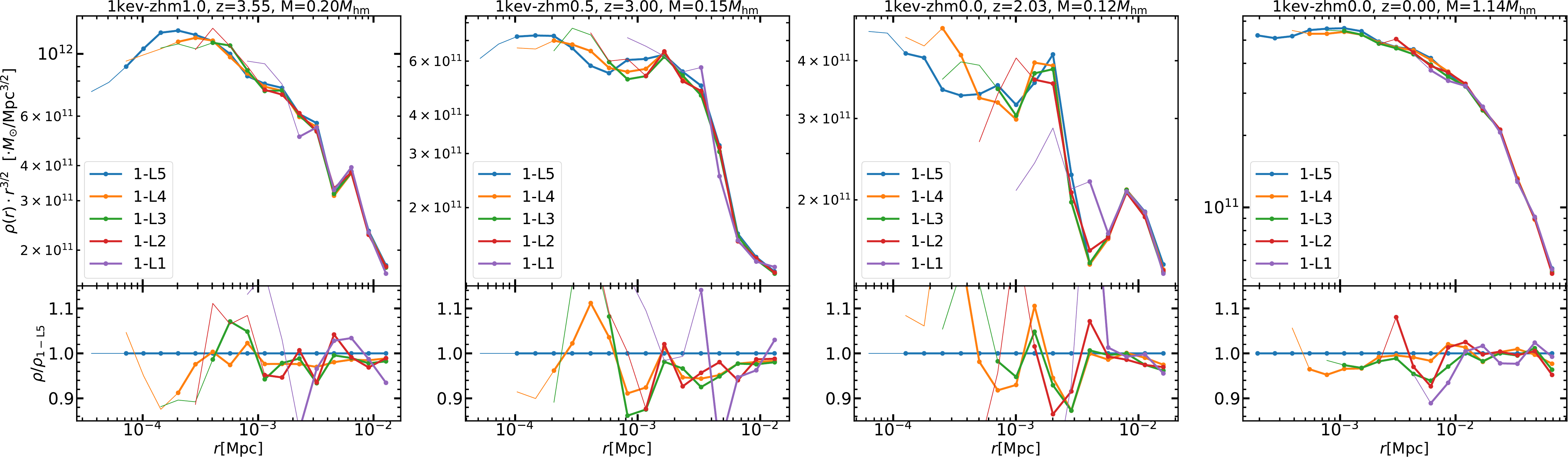}
    %{figures/warmsim1_obj5_sheetvsnb_density_convergence.pdf}
    \caption{Convergence of halo density profiles in zoom-in $N$-body simulations. In the upper row we display the density profile multiplied by $r^{1.5}$, and in the lower row the ratio with respect to the highest resolution. In each column we show a halo, where the colors depict different zoom levels (the numerical details are listed in Table \ref{tab:params_zooms}).}
    \label{fig:convergence_Nb}
\end{figure*}

In Figure \ref{fig:convergence_Nb} we show the radial structure of three first haloes for $m_{\rm DM}=1 \rm keV$ as given by $N$-body simulations: the density profiles multiplied by $r^{1.5}$ in the upper row, and the ratio with respect to the highest resolution simulation in the lower row. The different zoom-in levels are indicated in the legend, referring to Table \ref{tab:params_zooms}. In each zoom, we increase the resolution by a factor of 8. In total, from 1-L1 to 1-L5, the mass resolution of the simulations has increased by a factor of 4096. At the same time, the spatial resolution is 32 times better in 1-L5 than in 1-L1.

The first three columns show the haloes in some early stage, when they have acquired only $10-20\%$ of the half-mode mass. The agreement among different resolution levels is overall good. The outer part of the profiles are converged to few percent, while in the inner regions we get to $\sim 10\%$. In the last column we plot again the halo 1kev-zhm0.0, at $z=0.0$ with $M=1.14 M_{\rm hm}$. It is clear that, as the halo has grown, different zoom-in levels have converged to few percent in all the radial range. 

Hence, the early stages of halo formation and collapse are noisy, and hard to converge to good precision among different simulations -- even with exactly same initial conditions and gravity solver. It is important to keep this variability in mind when making comparisons at such early stages.

\subsection{Refinement + sheet}\label{sec:conv_sheet}
\begin{figure*}
    \centering
    \includegraphics[width=0.9\textwidth]{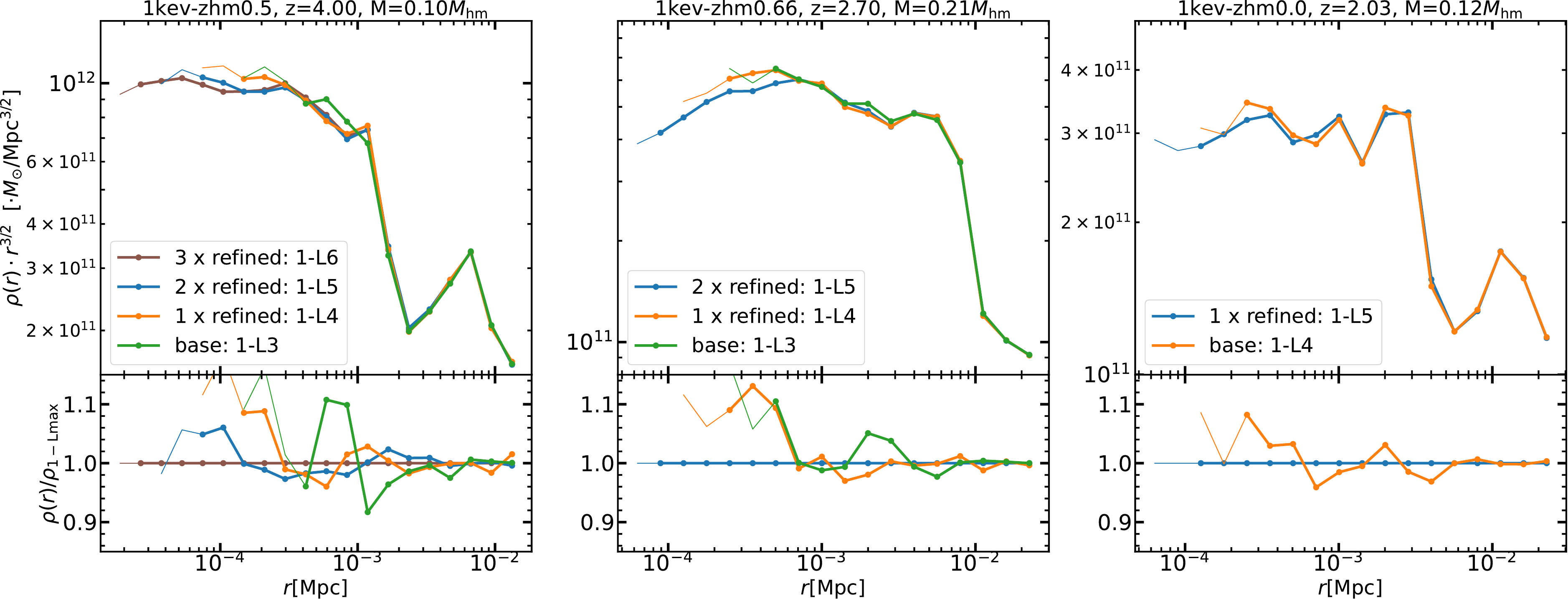}
    \caption{Convergence of halo density profiles in zoom-in refined S+R simulations. In the upper row we display the density profile multiplied by $r^{1.5}$, and in the lower row the ratio with respect to the highest resolution. In each column we show a halo, where the colors depict different resolution levels. We plot the base S+R simulation, and the refined simulations on top of it. The equivalent resolution of the simulations is listed in the legend, referring to Table \ref{tab:params_zooms}.}
    \label{fig:convergence_sh}
\end{figure*}

In this section we present the convergence of the refinement technique introduced in Section \ref{sec:refinement}. In Figure \ref{fig:convergence_sh} we display the halo density profiles of three haloes at different resolution levels, in a similar way to Figure \ref{fig:convergence_Nb} (check Section \ref{sec:conv_Nb} for details). The base simulation is the sheet-based zoom that we use as a reference to apply the refinement technique -- the lowest resolution line in the figure. In each refinement, we increase the mass resolution of the halo by a factor of eight, and the spatial resolution by a factor of two.

In the outer regions, the agreement between the base and refined simulations is remarkable. In the inner regions, similar to Figure \ref{fig:convergence_Nb}, they are converged to about $10\%$.

\subsection{With small-scale structure}\label{sec:conv_noise}
\begin{figure}
    \centering
    \includegraphics[width=0.8 \columnwidth]{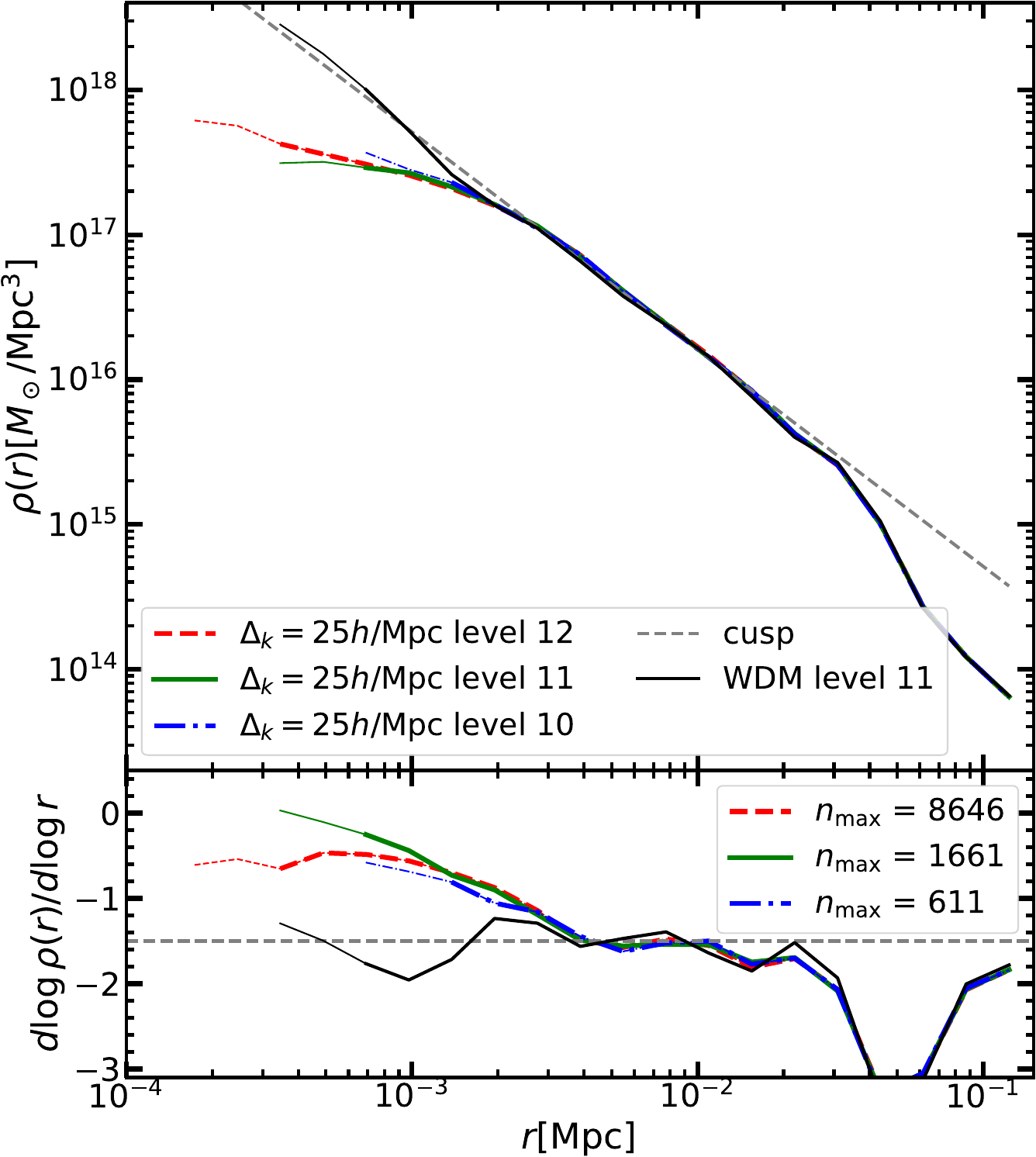}
    \caption{Resolution dependence of the density profiles of the haloes formed in presence of small scale power with $\Delta_k = 25 h/\rm Mpc$. In the upper panel we show the spherically averaged density profile ($\rho(r)$), and in the lower its logarithmic slope ($d \log \rho / d \log r$). Everything is given in physical units. The green, red and blue lines show the profiles measured in three identical simulations, varying only their resolution. We also display the number of particles that resolve the largest small-scale clumps in each case. For comparison, we plot the same halo in WDM cosmology at the middle resolution as a solid black line. }
    \label{fig:noise_dens_res}
\end{figure}

\begin{figure}
    \centering
    \includegraphics[width=0.8 \columnwidth]{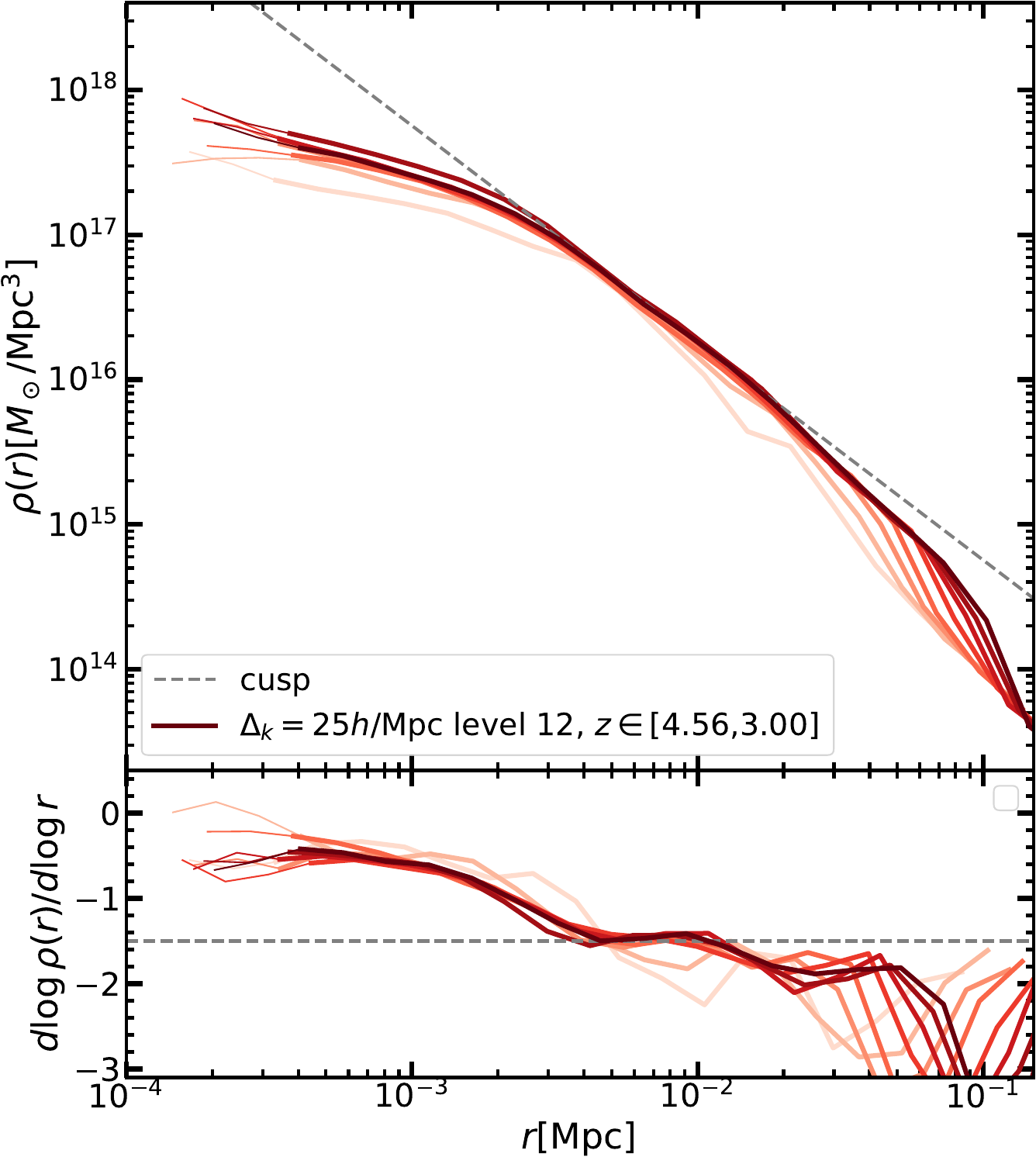}
    \caption{Redshift evolution of the density profile of the highest resolution halo of Figure \ref{fig:noise_dens}. In the upper panel we show the spherically averaged density profile ($\rho(r)$), and in the lower its logarithmic slope ($d \log \rho / d \log r$). Everything is given in physical units.}
    \label{fig:noise_dens_redshift}
\end{figure}

The presence of small-scale power in the collapse of a first halo can modify the shape of its inner structure: within a certain radius ($r_{\rm max}$), the profile becomes shallower and deviates from the prompt cusp that emerges in absence of small-scale power. In this section, we want to test the convergence and stability of this behaviour, both in regard to the position of $r_{\rm max}$, and to the shape of the profile in the inner region $r<r_{\rm max}$. 

In Figure \ref{fig:noise_dens_res} we show the density profiles of a halo. The black line depicts the halo that has formed in a pure WDM cosmology, without power at $k>k_{\rm hm}$. The colored lines are profiles for the halo formed with small-scale power for $\Delta_k = 25 h/\rm Mpc$, obtained from three simulations with different resolution. As indicated in the legend, in the blue case (the lowest resolution) the largest clumps are poorly resolved with $\sim 600$ particles, while in the red case (the highest resolution), the largest clumps have $\sim 8600$ particles. We first note that $r_{\rm max}$ seems to be well converged: at all resolutions, the colored lines deviate from the prompt cusp at $r_{\rm max} \sim 4 \rm kpc$. When we focus on the behaviour of the profile at $r<r_{\rm max}$, the colored lines differ slightly among themselves. For instance, in the green line the profile seems to be cored; while the red line shows a steeper slope. This means that the shape of the final halo at $r<r_{\rm max}$ depends on how well we resolve the small-scale clumps; the behaviour at $r<r_{\rm max}$ in Figure \ref{fig:noise_dens} is resolution-dependent.

In Figure \ref{fig:noise_dens_redshift} we show the redshift evolution of the density profiles of the highest resolution halo from  Figure \ref{fig:noise_dens}. It is clear that once the profile is set, there is little evolution in its inner regions. Importantly, the deviation point from the prompt cusp, $r_{\rm max}$, does not evolve with redshift, and the innermost region $r<r_{\rm max}$ appears to be stable. 

Summing up, the shape of the profiles at $r<r_{\rm max}$ depends on how well we resolve the small-scale clumps. More tests and higher resolution simulations would be needed to establish the shape of the profiles in this regime. However, the deviation point from the prompt cusp, $r_{\rm max}$, is stable with resolution and redshift, and it is converged even when the small clumps are resolved with a few hundred particles. 

\section{Pseudo phase-space density constraint}\label{sec:phase_space}

\begin{figure}
    \centering\includegraphics[width=0.8 \columnwidth]{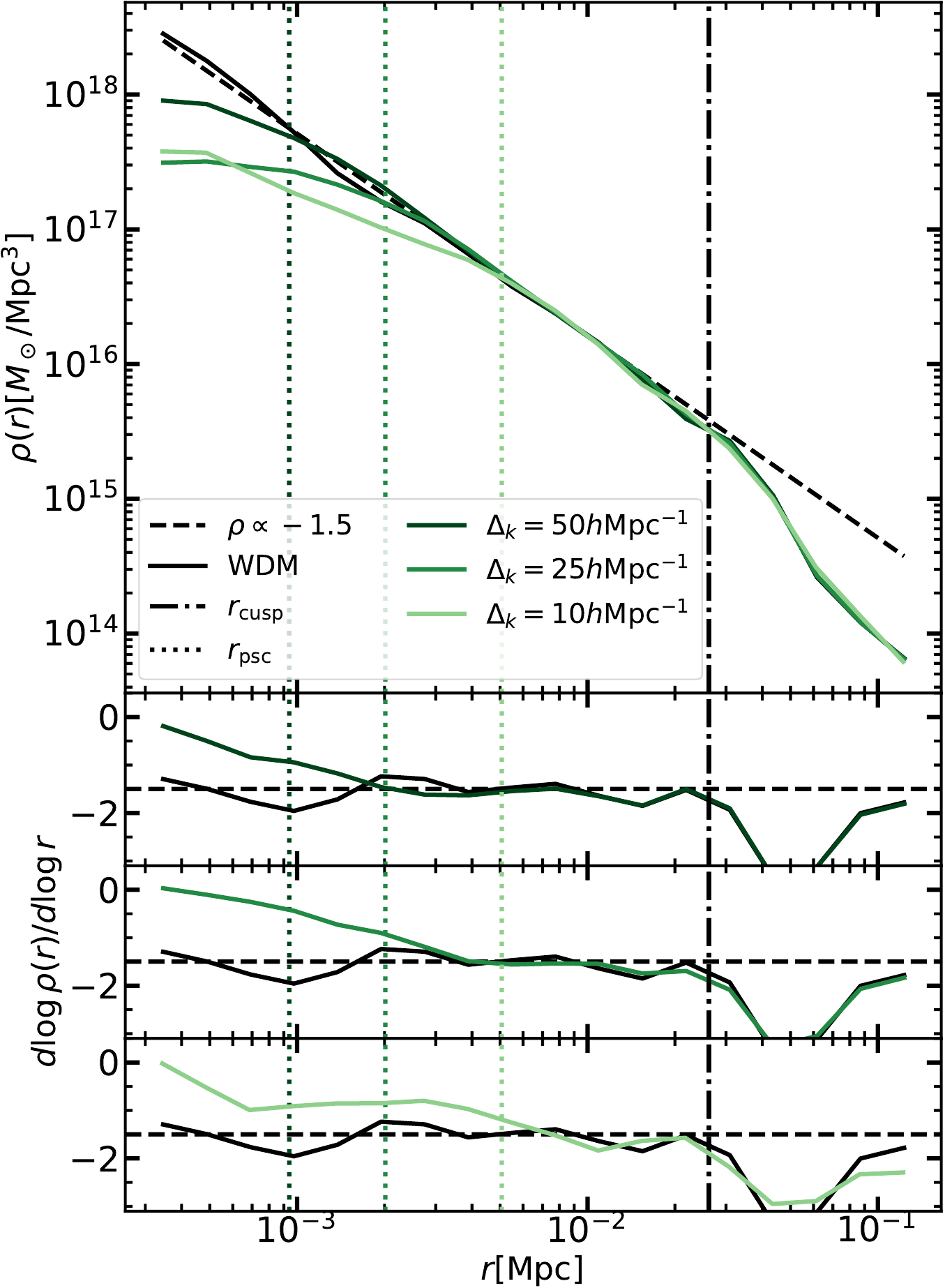}
    \caption{Density profiles of the haloes shown in Figure~\ref{fig:noise_warm2_obj3}, as in Figure \ref{fig:noise_dens}. The dotted lines denote $r_{\rm psc}$, the radii where the profiles deviate from the prompt cusp due to phase-space density constraints. For more details, see discussion in Appendix \ref{sec:phase_space}.}
    \label{fig:phs_constr}
\end{figure}

Dark matter haloes are expected to form thermal cores in their centers \citep{Tremaine:1979}. This is a consequence of Liouville's theorem, which states that the fine-grained phase-space density of dark matter is constant as a function of time. This value was set at dark matter freeze-out at the early universe, due to its non-zero velocity dispersion. The coarse-grained phase-space density of dark matter can never exceed this limit. Thus, when this value is saturated in halo centers, thermal cores are expected to form. 

In our experiments from Section \ref{sec:artfrags} a large halo forms from a distinct small scale medium which has already evolved small scale structures. For different scales of the small scale perturbations, the phase space distribution is heated to different degrees, prior to the formation of our main halo. This leads before collapse to a coarse-grained phase space density\footnote{Here coarse-grained means coarse-grained in velocity and in physical space.} that is significantly lower than in absence of the perturbations. We can expect that differences with respect to the fiducial profile occur at radii where the coarse-grained phase space density $f_{\rm{coarse}}$ is of order of the phase space density required to maintain the profile
\begin{align}
  f(\phi(r_{\rm{psc}})) = f_{\rm{coarse}}
\end{align}
This is the same constraint that is used calculate actual phase space cores in \cite{Delos:2019,Stucker:2023}, but applied to a coarse-grained phase space density. We call the corresponding radius the ``pseudo-core radius'' $r_{\rm{psc}}$ and we do not expect a uniform density for $r < r_{\rm{psc}}$, since phase space densities may still be larger than this coarse grained value in some regions. However, the small scale noise will limit much of the material to $f \lesssim  f_{\rm{coarse}}$ so that we may expect a deviation from the $r^{-1.5}$ powerlaw behavior below $r_{\rm{psc}}$.

We can estimate $f_{\rm{coarse}}$ from the expected small scale motion under the Zel'dovich approximation at the time of collapse of the halo. The peculiar velocity of dark matter at at time $a$ is under the Zel'dovich approximation
\begin{align}
    \Vec{v} &= a f(a) H(a) D(a) \Vec{s}_0
\end{align}
where $\Vec{s}_0$ is the displacement field at $a=1$, $D$ is the linear growth factor normalized to $D(a=1)=1$ and $f = \mathrm{d} \log D / \mathrm{d} \log a \approx \Omega_m(a)^{5/9}$ \citep[e.g. ][]{Jenkins:2010}. Therefore, the peculiar velocity dispersion induced by gravity is given by
\begin{align}
    \sigma_v(a) &= a f(a) H(a) D(a) \sigma_{s,0}
\end{align}
where $\sigma_{s,0}$ is the variance of the displacement field at $a=1$ which can be calculated as 
\begin{equation}
    \sigma_{s,0} = \sqrt{\int {\frac{P(k)}{2 \pi^2} dk}}
\end{equation}
If we use for $P(k)$ the noise spectrum $P(k) = T_{\rm noise}(k)P_{\rm CDM}(k)$ (see Equation \ref{eq:noise_transfer}), then $\sigma_v(a)$ reflects the degree of random motion that we expect from the small scale medium at time $a$. Since for a larger halo made of this medium, the internal perturbations can only grow up to the collapse time $a_{\rm col}$, an estimate of the coarse-grained phase space density constraint imposed to the halo is given by
\begin{align}
    f_{\rm{coarse}}  = \frac{1}{(2\pi)^{3/2}} \frac{\rho_{m}(a_{\rm{col}})} {\sigma_v(a_{\rm{col}})^3},
\end{align}
where $\rho_m(a_{\rm col})$ is the background density of the Universe at collapse time. Following \cite{Stucker:2023,Delos:2019}, this implies 
\begin{equation}\label{eq:rcore}
    r_{\rm psc} = \frac{3}{64} \left[\frac{3 \cdot 70^{4}}{\pi^{10}}\right]^{1/9} G^{-2/3} A^{-2/9}f_{\rm coarse}^{-4/9},
\end{equation}
where $A$ is the amplitude of the profile and $G$ is the universal gravitational constant.

In Figure \ref{fig:phs_constr} we show the density profiles of a halo forming in three mediums with different degree of small scale noise (see Section \ref{sec:small-scale-noise} and Figure \ref{fig:noise_dens} for a detailed description). The vertical dotted lines denote the prediction for the "pseudo-core". It is clear that the radii derived from the phase-space density limit fall where the profiles start deviating from a pure power-law solution. However, note that at $r<r_{\rm psc}$ the profiles do not necessarily form cores.

% Don't change these lines
\bsp	% typesetting comment
\label{lastpage}
\end{document}